\documentclass[12pt]{article}

\usepackage{mathrsfs}
\usepackage[T1]{fontenc}
\usepackage{mathpazo}
\usepackage{setspace}
\usepackage{amsfonts}
\usepackage{amssymb}
\usepackage{amsmath}
\usepackage{esint}                   
\usepackage{epsfig}
\usepackage{latexsym}
\usepackage{color}
\usepackage{graphicx}
\usepackage{hyperref}
\usepackage{nicefrac}
\usepackage[latin1]{inputenc}
\usepackage{pstricks}
\usepackage{slashed}
\usepackage{multirow}
\usepackage{ulem}
\usepackage{comment}
\usepackage[nosort]{cite}
\usepackage{datetime}
\usepackage{tikz-cd}
\usepackage{float}

\usepackage{braket}

\usepackage[greek,english]{babel}
\languageattribute{greek}{polutoniko}

\def\hybrid{\topmargin -30pt    \oddsidemargin 0pt 
        \headheight 0pt \headsep 0pt
        \textwidth 6.25in       
        \textheight 9.5in       
        \marginparwidth .875in
        \parskip 5pt plus 1pt   \jot = 1.5ex}

\hybrid

\def\baselinestretch{1.2}

\catcode`\@=11

\def\marginnote#1{}
\newcount\hour
\newcount\minute
\newtoks\amorpm
\hour=\time\divide\hour by60
\minute=\time{\multiply\hour by60 \global\advance\minute by-\hour}
\edef\standardtime{{\ifnum\hour<12 \global\amorpm={am}%
        \else\global\amorpm={pm}\advance\hour by-12 \fi
        \ifnum\hour=0 \hour=12 \fi
        \number\hour:\ifnum\minute<10 0\fi\number\minute\the\amorpm}}
\edef\militarytime{\number\hour:\ifnum\minute<10 0\fi\number\minute}

\def\draftlabel#1{{\@bsphack\if@filesw {\let\thepage\relax
   \xdef\@gtempa{\write\@auxout{\string
      \newlabel{#1}{{\@currentlabel}{\thepage}}}}}\@gtempa
   \if@nobreak \ifvmode\nobreak\fi\fi\fi\@esphack}
        \gdef\@eqnlabel{#1}}
\def\@eqnlabel{}
\def\@vacuum{}
\def\draftmarginnote#1{\marginpar{\raggedright\scriptsize\tt#1}}

\def\draft{\oddsidemargin -.5truein
        \def\@oddfoot{\sl preliminary draft \hfil
        \rm\thepage\hfil\sl\today\quad\militarytime}
        \let\@evenfoot\@oddfoot \overfullrule 3pt
        \let\label=\draftlabel
        \let\marginnote=\draftmarginnote
   \def\@eqnnum{(\theequation)\rlap{\kern\marginparsep\tt\@eqnlabel}%
\global\let\@eqnlabel\@vacuum}  }

\def\draft2{
        \def\@oddfoot{\sl preliminary draft \hfil
        \rm\thepage\hfil\sl\today\quad\militarytime}
        \let\@evenfoot\@oddfoot \overfullrule 3pt
        \let\marginnote=\draftmarginnote
   \def\@eqnnum{(\theequation)\rlap{\kern\marginparsep\tt\@eqnlabel}%
\global\let\@eqnlabel\@vacuum}  }

\def\preprint{\twocolumn\sloppy\flushbottom\parindent 2em
        \leftmargini 2em\leftmarginv .5em\leftmarginvi .5em
        \oddsidemargin -.5in    \evensidemargin -.5in
        \columnsep .4in \footheight 0pt
        \textwidth 10.in        \topmargin  -.4in
        \headheight 12pt \topskip .4in
        \textheight 6.9in \footskip 0pt
        \def\@oddhead{\thepage\hfil\addtocounter{page}{1}\thepage}
        \let\@evenhead\@oddhead \def\@oddfoot{} \def\@evenfoot{} }

\def\numberbysection{\@addtoreset{equation}{section}
        \def\theequation{\thesection.\arabic{equation}}}

\def\underline#1{\relax\ifmmode\@@underline#1\else
        $\@@underline{\hbox{#1}}$\relax\fi}

\def\titlepage{\@restonecolfalse\if@twocolumn\@restonecoltrue\onecolumn
     \else \newpage \fi \thispagestyle{empty}\c@page\z@
        \def\thefootnote{\fnsymbol{footnote}} }

\def\endtitlepage{\if@restonecol\twocolumn \else \newpage \fi
        \def\thefootnote{\arabic{footnote}}
        \setcounter{footnote}{0}}  

\catcode`@=12
\relax

\def\figcap{\section*{Figure Captions\markboth
        {FIGURECAPTIONS}{FIGURECAPTIONS}}\list
        {Figure \arabic{enumi}:\hfill}{\settowidth\labelwidth{Figure
999:}
        \leftmargin\labelwidth
        \advance\leftmargin\labelsep\usecounter{enumi}}}
 \relax
\def\tablecap{\section*{Table Captions\markboth
        {TABLECAPTIONS}{TABLECAPTIONS}}\list
        {Table \arabic{enumi}:\hfill}{\settowidth\labelwidth{Table
999:}
        \leftmargin\labelwidth
        \advance\leftmargin\labelsep\usecounter{enumi}}}
 \relax
\def\reflist{\section*{References\markboth
        {REFLIST}{REFLIST}}\list
        {[\arabic{enumi}]\hfill}{\settowidth\labelwidth{[999]}
        \leftmargin\labelwidth
        \advance\leftmargin\labelsep\usecounter{enumi}}}
 \relax
\makeatletter
\newcounter{pubctr}
\def\publist{\@ifnextchar[{\@publist}{\@@publist}}
\def\@publist[#1]{\list
        {[\arabic{pubctr}]\hfill}{\settowidth\labelwidth{[999]}
        \leftmargin\labelwidth
        \advance\leftmargin\labelsep
        \@nmbrlisttrue\def\@listctr{pubctr}
        \setcounter{pubctr}{#1}\addtocounter{pubctr}{-1}}}
\def\@@publist{\list
        {[\arabic{pubctr}]\hfill}{\settowidth\labelwidth{[999]}
        \leftmargin\labelwidth
        \advance\leftmargin\labelsep
        \@nmbrlisttrue\def\@listctr{pubctr}}}
 \relax
\makeatother

\def\be{\begin{equation}}
\def\ee{\end{equation}}
\def\ba{\begin{eqnarray}}
\def\ea{\end{eqnarray}}

\def\del{\partial}

\def\a{\alpha}

\def\b{\beta}

\def\d{\delta}
\def\D{\Delta}

\def\l{\lambda}

\def\no{\noindent}

\def\qq{\qquad}

\def\IR{\relax{\rm I\kern-.18em R}}

\def\inv{^{\raise.0ex\hbox{${\scriptscriptstyle -}$}\kern-.05em 1}}

\def \ha {{\frac{1}{2}}}

\def \ov {\over}

\newcommand{\bb}{\hskip -0.1cm}

\def\tr{\textrm{Tr}}

\begin{document}


\renewcommand{\theequation}{\thesection.\arabic{equation}}
\csname @addtoreset\endcsname{equation}{section}

\begin{titlepage}
\begin{center}

\hfill CERN-TH-2023-086

\renewcommand*{\thefootnote}{\arabic{footnote}}

\phantom{xx}
\vskip 0.4in

{\large {\bf Ferromagnetic phase transitions in $SU(N)$ }}

\vskip 0.4in

{\bf Alexios P. Polychronakos$^{1,2}$}\hskip .15cm and \hskip .15cm
{\bf Konstantinos Sfetsos}$^{3,4}$

\vskip 0.14in

${}^1\!$ Physics Department, the City College of the New York\\
160 Convent Avenue, New York, NY 10031, USA\\
{\footnotesize{\tt apolychronakos@ccny.cuny.edu}}\\
\vskip 0.3cm
${}^2\!$ The Graduate School and University Center, City University of New York\\
365 Fifth Avenue, New York, NY 10016, USA\\
{\footnotesize{\tt apolychronakos@gc.cuny.edu}}

\vskip .14in

${}^3\!$
Department of Nuclear and Particle Physics, \\
Faculty of Physics, National and Kapodistrian University of Athens, \\
Athens 15784, Greece\\
{\footnotesize{\tt ksfetsos@phys.uoa.gr}}\\

\vskip .14in
${}^4\!$ Theoretical Physics Department,
 CERN, \\1211 Geneva 23, Switzerland

\vskip .3in
\today

\vskip .2in

\end{center}

\vskip .2in

\centerline{\bf Abstract}

\no
We study the thermodynamics of a non-abelian ferromagnet consisting of "atoms"
each carrying a fundamental representation of $SU(N)$, coupled with long-range two-body
quadratic interactions. We uncover a rich structure of phase transitions from non-magnetized,
global $SU(N)$-invariant states to magnetized ones breaking global invariance to $SU(N-1) \times U(1)$.
Phases can coexist, one being stable and the other metastable, and the transition between states involves
latent heat exchange, unlike in usual $SU(2)$ ferromagnets. Coupling the system to an
external non-abelian magnetic field further enriches the phase structure, leading to additional phases.
The system manifests hysteresis phenomena both in the magnetic field, as in usual ferromagnets,
and in the temperature, in analogy to supercooled water. Potential applications are in fundamental
situations or as a phenomenological model.

\vskip .4in

\vfill

\end{titlepage}
\vfill
\eject



\def\baselinestretch{1.2}
\baselineskip 20 pt

\newcommand{\eqn}[1]{(\ref{#1})}

\tableofcontents


\section{Introduction}
\label{intro}

Magnetic materials are of considerable physical and technological interest, and their properties have
long been the subject of theoretical research. Ferromagnets, the first type of magnetism ever observed,
hold a special place among them, as they manifest nontrivial properties and symmetry breaking.

All known ferromagnets consist of interacting localized magnetic dipoles and break rotational invariance
below the Curie temperature. Each (quantum) dipole provides a representation of the group of rotations,
that is, of $SU(2)$. Although this is a nonabelian group, it is of a particularly simple type: it has a
unique Cartan generator, and $SU(2)$ dipoles can interact with external abelian magnetic fields that
couple to their Cartan generator. Nevertheless, the exact quantitative properties of physical ferromagnets
remain an active topic of research \cite{LGF}.

Independently, nonabelian unitary groups $SU(N)$ of higher rank play a crucial role in particle physics
and, indirectly through matrix models, in string theory and gravity. Ungauged and gauged $SU(3)$ groups
are the most common, representing "flavor" and "color" degrees of freedom, respectively.
A collection of nucleons,
or the constituents of the quark-gluon plasma, are physical systems of components carrying
representations of $SU(3)$. This raises the obvious question of the properties that large collections of
such $SU(3)$ or, more generally, $SU(N)$ entities would have if they interacted with each other as
well as with external nonabelian magnetic fields.

In this work we investigate the properties of such systems in the ferromagnetic regime, that is,
in the regime where the mutual interaction of its components would tend to "align" their $SU(N)$
charges, in a way that we will make precise. The results on the decomposition of the direct
product of an arbitrary number of representations of $SU(N)$ into irreducible components that we
derived in a recent publication \cite{SUN} will be a crucial tool in our calculations.
We will also study the effects
of an external nonabelian magnetic field coupled to the system. As we shall demonstrate,
the properties of nonabelian ($N>2$) ferromagnets are qualitatively different from these of ordinary
($N=2$) ferromagnets. They display a rich phase structure involving various critical temperatures,
hysteresis both in the temperature and in the magnetic field, coexistence of phases, and latent heat
transfer during phase transitions.

In the sequel we will present the basic $SU(N)$ model, consisting of distinguishable quantum
components in the fundamental representation, and will review the relevant group theory results of
\cite{SUN}.
We will proceed to study the thermodynamic phases of the model in the absence, and subsequently in the
presence, of external magnetic fields, and will derive its symmetry breaking patterns, critical
temperatures, and magnetization. Stability issues will be crucial and will determine the pattern of
$SU(N)$ breaking in the various phases. We will further study the nontrivial situation
of a magnetic field inducing an enhanced breaking of $SU(N)$, and will conclude with some speculations
about the phenomenological relevance of the model.


\section{A system of interacting $SU(N)$ "atoms"}
\label{rev&model}

Magnetic systems with $SU(N)$ symmetry have been considered in the context of ultracold atoms
\cite{Ghu,Gor,Zha,Mag,Cap} or of interacting atoms on lattice cites \cite{KT,BSL,RoLa,YSMOF,TK,Totsuka,TK2}
and in the presence of $SU(N)$ magnetic fields \cite{DY,YM,HM}.

In this section we lay out the
basic structure of any model of interacting $SU(N)$ atoms, specify its ferromagnetic regime,
and review the group theory results necessary for its analytic treatment.

 \subsection{The model}

To motivate the basic model, consider a set of $n$ atoms (or molecules) on a lattice, interacting with
two-body interactions. Each atom is in one of $N$ degenerate or
quasi-degenerate states $\ket{s}$, $s = 1,2, \dots, N$. The generic two-body interaction between atoms
$1$ and $2$ with states $\ket {s_1}$ and $\ket {s_2}$ would be
\be
H_{12} = \sum_{s_1 , s_1' , s_2 , s_2' = 1}^N h_{s_1 s_2 ; s_1' s_2'} \ket{s_1} \bra{s_1'} \otimes 
\ket{s_2} \bra{s_2'} \ ,\qq h_{s_1 s_2 ; s_1' s_2'} = h_{s_1' s_2' ; s_1 s_2}^*\ .
\ee
Define $j_a$, $a=0,1,\dots,N^2 -1$, the generators of $U(N)$ in the fundamental $N$-dimensional
representation, with $j_0$ the identity operator (the $U(1)$ part). Using the fact that the $j_a$ form a
complete basis
for the operators acting on an $N$-dimensional space, the above interaction can also be written as
\be
H_{12} = \sum_{a , b = 0}^{N^2 -1} h_{a b} \, j_{1,a} \, j_{2,b} \ ,\qq h_{ab} = h_{ab}^*\ ,
\label{hab}\ee
where
\be 
j_{1,a} = j_a \otimes 1 \ ,\qq j_{2,a} = 1 \otimes j_a\ ,
\ee
are the fundamental $U(N)$ operators acting on the states of atoms $1$ and $2$. 

We now make the physical assumption that the above interaction is invariant under a change of basis in
the states $\ket{s}$, that is, under a common unitary transformation of the states $\ket{s_1}$ of atom
$1$ and $\ket{s_2}$ of atom $2$. This implies two equivalent facts:
first, the interaction will necessarily be, up to trivial additive and multiplicative constants
(proportional to the identity), the operator exchanging the states of the atoms,
\be
H_{12} = C_{12}' + C_{12} \sum_{s , s' = 1}^N \ket{s} \bra{s'} \otimes \ket{s'} \bra{s}\ ,
\ee
with $C_{12}' , C_{12}$ being real constants. 
Second, the interaction will necessarily be of the form
\be
H_{12} = c_{12}' + c_{12} \sum_{a = 1}^{N^2 -1} j_{1,a} \, j_{2,a}\ .
\ee
Omitting the trivial constant $c_{12}'$ (due to the $U(1)$ part), we obtain a unique two-body
interaction depending on a single coupling constant $c_{12}$.
Note that the group $U(N)$ emerges from the requirement of invariance under general changes of basis
of the $N$ states, and leads to interactions linear in the
operators in each atom. Using, instead, an $N$-dimensional representation of a smaller group would
require the inclusion of higher polynomial terms in $j_{1,a}$ and $j_{2,a}$. 

The interaction Hamiltonian of the full system will be of the form
\be
H = \sum_{r,s =1}^n c_{r s} \sum_{a= 1}^{N^2 -1} j_{r,a} \, j_{s,a}\ ,
\ee
where $c_{rs} = c_{sr}$ is the strength of the interaction between atoms $r$ and $s$ (and $c_{rr}=0$).
This Hamiltonian involves an isotropic quadratic coupling between the fundamental generators of
the $n$ commuting $SU(N)$ groups of the atoms. 

Reasonable physical assumptions restrict the form of the couplings $c_{rs}$. We assume that the
interaction is homogeneous, that is, $c_{rs}$ is translationally invariant under the shift of both $r$
and $s$ by the same lattice translation (away from the boundary of the lattice).
In terms of the lattice positions of the atoms $\vec r$,
\be
c_{{\vec r} , {\vec s}} = c_{{\vec r} - {\vec s}} \ ,\qq c_0 = 0\ .
\ee
Therefore, each atom couples to a fixed weighted average of the $SU(N)$ generators of its neighboring
atoms. We will also assume that interactions are reasonably long-range, that is, each atom couples to
several of its neigboring atoms. This technical assumption justifies the mean field condition that,
in the thermodynamic limit,
the weighted average of the neighboring atoms is well approximated by their average over the full
lattice.  That is
\be
\sum_{\vec s} c_{\vec s}\, j_{{\vec r}+{\vec s},a} \simeq \Bigl(\sum_{{\vec s}} c_{\vec s} \Bigr)\,
{1\over n} \sum_{s=1}^n j_{s,a} = -{c \over n} J_a\ ,
\label{meanfield}\ee
where we defined the total $SU(N)$ generator
\be
J_a = \sum_{s=1}^n j_{s,a}
\ee
and the effective mean coupling\footnote{The validity of the mean field approximation is
strongest in three dimensions, since every atom has a higher number of near neighbors and the statistical
fluctuations of their averaged coupling are weaker, but is expected to hold also in lower dimensions.}
\be
c = - \sum_{{\vec s}} c_{\vec s}\ .
\ee
The minus sign is introduced such that ferromagnetic interactions, driving atom states to align,
correspond to positive $c$.
Altogether, the full effective interaction assumes the form
\be
H = -{c \over n} \sum_{a=1}^{N^2 -1}\Bigl(  J_a^2 - \sum_{s=1}^n j_{s,a}^2 \Bigr)\ ,
\ee
where the second term in the parenthesis eliminates the terms $r=s$.
The first part of $H$ is proportional to
the quadratic Casimir of the total $SU(N)$ group $C^{(2)} = \sum_a J_a^2$. The second part is
proportional to the sum of the quadratic Casimirs of each individual atom. Since all $j_{s,a}$ are in
the fundamental representation, their quadratic Casimir is a (common) constant, independent of their
state. So the second term contributes a trivial constant and can be discarded.

In addition to the atoms' mutual interaction, we can couple the states of the atoms to a global external
field, contributing an additional term
\be
H_B = \sum_{r=1}^n \sum_{s , s' =1}^N B_{s s'} \ket{s}_r \bra {s'}_r\ .
\ee
Parametrizing this one-atom operator in terms of the complete set of operators $j_{r,a}$ and
omitting the trivial constant terms corresponding to $j_{r,0}$, it becomes
\be
H_B = \sum_{r=1}^n \sum_{a=1}^{N^2 -1} B_a \, j_{r,a} = \sum_{a=1}^{N^2 -1} B_a J_a\ .
\ee
We see that $B_a$ acts as a global nonabelian magnetic field on the $SU(N)$ "spins" $j_{r,a}$.
Finally, making use of the fact that the interaction Hamiltonian is invariant under global $SU(N)$
transformations, we may choose a basis of states in which the sum $\sum_a B_a J_a$
is rotated to the Cartan subspace spanned by the commuting generators $H_i$, $i=1,2,\dots,N-1$.
The full Hamiltonian of the model then emerges as
\be
H = -{c\ov n} C^{(2)} - \sum_{i=1}^{N-1} B_i H_i\ .
\ee
We will assume that $c$ is positive, so that the model is of the ferromagnetic type.

For $N=2$ the above model reduces to the ferromagnetic interaction of spin-half components.
For higher $N$, the model has the same number of states per atom as a spin-$S$ $SU(2)$
model with $2S+1 = N$. The dynamics of the two models, however, are distinct: the $SU(2)$ model
is invariant only under global $SU(2)$ transformations, which cannot mix the $N$ states of the atoms
in an arbitrary way, unlike the $SU(N)$ case.
The enhanced symmetry of the $SU(N)$ model leads, as we shall see, to a richer structure and
to qualitatively different thermodynamic properties.

Finding the eigenstates of the above model and determining its thermodynamics involves
decomposing the full Hilbert space of states into irreducible representations (irreps) of the total $SU(N)$,
evaluating the quadratic Casimir $C^{(2)}$ and the magnetic sum $\sum_i B_i H_i$ in each irrep,
and calculating the partition function as a sum over these irreps. This requires determining
the decomposition of the direct product of a large, arbitrary number $n$ of $SU(N)$ fundamentals
into irreps and the multiplicity of each irrep in the decomposition, as well as calculating the Casimir
and the magnetic sum for large irreps of $SU(N)$. This task was performed in a
recent publication \cite{SUN}, and the relevant results will be reviewed in the next subsection.

\subsection{Decomposition of $n$ fundamentals of $SU(N)$ into irreps}

We summarize the group theory results pertaining to the decomposition of the direct product of $n$
fundamentals of $SU(N)$ into irreps, as presented in \cite{SUN} (results on the simpler case of $SU(2)$
were previously derived in \cite{SU2,Zax} and were applied in \cite{SU2} to regular ferromagnetism).

The setting and results become most
tractable and intuitive in the momentum representation, in which irreps of $SU(N)$ are
labeled by a set of distinct integers $k_i$, $i=1,2,\dots ,N$ ordered as
\be
\label{k1k2}
k_1 > k_2 > \dots > k_N\ .
\ee
Each irrep corresponds to a given set $\{k_i\}$, for which we will use the symbol $\bf k$.
The corresponding Young Tableaux (YT) of the irrep may be described by its lengths $\ell_i$,
i.e., number of boxes per row, for $i=1,2,\dots ,N-1$. The correspondence with $k_i$ is
\be
\label{ellk}
\ell_i = k_i - k_N + i-N\ , \qq \ell_1 \geqslant \ell_2\geqslant\dots \geqslant \ell_{N-1} \geqslant 0\ .
\ee
Note that the $k_i$ representation is redundant, since a shift of all $k_i$ by a common
constant $k_i \to k_i + c$ leaves $\ell_i$ invariant and leads to the same irrep of $SU(N)$
(the shift changes the $U(1)$ charge of the irrep, which equals the sum of the $k_i$).
This freedom can be used to simplify relevant formulae. In our situation, where irreps will arise from
the direct product of $n$ fundamentals, it will be convenient to choose the convention
\be
\sum_{i=1}^N k_i = n +{N(N-1)\over 2}\ .
\label{suma}
\ee
For the singlet representation ($n=0$) all $\ell_i$ are zero, which in the above convention corresponds
to $k_i=N-i$, $i=1,2,\dots, N$. The fundamental ($n=1$) has a single box, and corresponds
to $k_1= N$ and the rest of the $k_i$ as above.

In $SU(N)$ there are $N-1$ Casimir operators which, for the irrep $\bf k$, can be expressed in terms of
the $k_i$'s. For our purposes we need the quadratic Casimir, which is given in terms of the $k_i$ by
\be
\label{Casimi}
C^{(2)} ({\bf k}) = {1\over 2} \sum_{i=1}^N k_i^2 - {1\over 2N} \left[n+N(N-1)/2\right]^2 
-{N(N^2 -1)\over 24} \ .
\ee
Note that, using (\ref{ellk}), (\ref{suma}), $C^{(2)}$ takes the more familiar form
\be
\label{Casimil}
C^{(2)} ({\bf \ell}) = {1\over 2} \sum_{i=1}^{N-1} \ell_i (\ell_i +N+1-2i) - {1\over 2N} 
\left(\sum_{i=1}^{N-1} \ell_i \right)^2  \ .
\ee
For the singlet $C^{(2)} = 0$, while for the fundamental $C^{(2)} = (N-N^{-1})/2$.

For our purposes we also need the trace of the exponential of the magnetic term in a giver irrep $\bf k$,
\be
\tr_{\bf k} \exp\Bigl( \beta \sum_{j=1}^N B_j H_j  \Bigr)\ ,
\ee
which will appear in the calculation of the partition function of our model.
This was calculated in \cite{SUN}. To express it, define the Slater determinant
\be
\label{slat}
\psi_{\bf k}({\bf z} ) = (z_1 \cdots z_N )^{-{1\over N}\sum_i k_i} \left|
\begin{array}{ccccccccc}
z_1^{k_1} & z_1^{k_2} & \cdots & z_1^{k_{N-1}}  & z_1^{k_N} \\
z_2^{k_1} & z_2^{k_2} & \cdots & z_2^{k_{N-1}}  & z_2^{k_N} \\
\vdots & \vdots & \ddots & \vdots & \vdots  \\
z_N^{k_1} & z_N^{k_2} & \cdots & z_N^{k_{N-1}} & z_N^{k_N}
\end{array}
\right |\ ,\qq {\bf z}=\{z_i \in \mathbb{C}\}\ ,
\ee
which is antisymmetric under the interchange of any two $z_i$'s and of any two $k_i$'s. Also define
the Vandermonde determinant
\be
\label{vanderm}
\Delta ({\bf  z}) = (z_1 \cdots z_N )^{-{N-1 \over 2}} \left|
\begin{array}{ccccccccc}
z_1^{N-1} & z_1^{N-2} & \cdots & z_1  & 1 \\
z_2^{N-1} & z_2^{N-2} & \cdots & z_2  & 1 \\
\vdots & \vdots & \ddots & \vdots & \vdots  \\
z_N^{N-1} & z_N^{N-2} & \cdots & z_N  & 1 \\
\end{array}
\right |\ ,
\ee
which is the Slater determinant (\ref{slat}) for the singlet irrep. Then
\be
\tr_{\bf k} \exp\Bigl( \beta \sum_{j=1}^N B_j H_j \Bigr) = {\psi_{\bf k}({ \bf z}) \over \Delta ({\bf z})}
\ , \qq z_j = e^{\beta B_j}\ .
\label{magtr}\ee
The prefactors involving the product $z_1 \cdots z_N$ in (\ref{slat}) and (\ref{vanderm}) eliminate the
$U(1)$ part of the irrep, which couples to the trace of the magnetic field $\sum_i B_i$. If  $B$ is
traceless, then the $U(1)$ charge decouples and we can ignore these prefactors.
As a check of (\ref{magtr}), we can take the limit $z_i \to 1$ and verify that the ratio of determinants goes
to
\be
\tr_{\bf k} {\bf 1} = \dim ({\bf k}) = \prod_{j>i=1}^N {k_i - k_j  \over j-i}
= \prod_{j>i=1}^N {\ell_i - \ell_j +j-i \over j-i}\ ,
\ee
which is the standard expression for the dimension of the irrep.

The last nontrivial element needed for our purposes is the multiplicity $d_{n,{\bf k}}$ of each irrep
$\bf k$ arising in the decomposition of $n$  fundamental representations. This was also calculated
in \cite{SUN}, and the result is
\be
\label{denn}
\begin{split}
d_{n, {\bf k}} &= \delta_{k_1 + \cdots +k_N ,n+N(N-1)/2} \prod_{j>i=1}^N (S_i - S_j ) D_{n, {\bf k}} \ ,
\\
D_{n, {\bf k}}  &= {n! \over \prod_{r=1}^N k_r !}\ ,
\end{split}
\ee
where $S_i$ is a shift operator acting on the right by replacing $k_i$ by $k_i -1$.
Note that $D_{n,{\bf k}}$ and $d_{n,{\bf k}}$ are manifestly symmetric and antisymmetric, respectively,
under exchange of the $k_i$. In \cite{SUN} the action of the operator $\prod_{j>i=1}^N (S_i - S_j )$ on
$D_{n, {\bf k}}$ was performed and an explicit combinatorial formula for $d_{n, {\bf k}}$ was obtained,
but it will not be needed for our purposes.

\subsection{The thermodynamic limit of the model}

We now have all the ingredients to study the statistical mechanics of our $SU(N)$ ferromaget.
The partition function is
\be
Z= \sum_{\text{states}} e^{-\beta H} =
\sum_{\braket{\bf k}} d_{n;{\bf k}}\, e^{{\beta c\over n} C^{(2)} ({\bf k})}\,
\tr_{\bf k} \exp\Bigl( \beta \sum_{j=1}^N B_j H_j  \Bigr)
\ ,
\ee
where $\b$ is the inverse of the temperature $T$ and $\braket{\bf k}$
denotes distinct ordered integers $k_1 > k_2 > \cdots >k_N$ satisfying the constraint \eqn{suma}.
Using the results (\ref{Casimi}), (\ref{magtr}) and (\ref{denn}), and removing
the trivial ($k_i$-independent) terms in the Casimir \eqn{Casimi}, the partition function becomes
\be
\begin{split}
Z
= &{1\over N!}\sum_{\bf k} \delta_{k_1 + \cdots +k_N ,n+{N(N-1)\over 2}}
\bigg(\prod_{j>i=1}^N (S_i - S_j ) 
{ n!  \over \prod_{r=1}^N k_r !}\bigg)~
{\psi_{\bf k} ({\bf z} ) \over \Delta ({\bf z})} \, e^{{\beta c \over 2n} \sum_s k_s^2}
\\
=& \sum_{\bf k} \delta_{k_1 + \cdots +k_N ,n+{N(N-1)\over 2}}\,
 {1\over\Delta ({ \bf z})} \bigg(\prod_{j>i=1}^N (S_i - S_j ) 
{n! \over \prod_{r=1}^N k_r !}\bigg) \,  
e^{{\b c \over 2n} \sum_s k_s^2 +\beta B_s k_s}
\\
=& \sum_{\bf k} \delta_{k_1 + \cdots +k_N ,n}\,
 {1\over \Delta ({ \bf z})}
{n! \over \prod_{r=1}^N k_r !} \,  
\prod_{j>i=1}^N \bigl(S_i^{-1} - S_j^{-1} \bigr)\, e^{{\b c \over 2n} \sum_s k_s^2 +\beta B_s k_s}\ .
\end{split}
\ee
In the first line above we made the sum unrestricted, since the summand is symmetric under
permutation of the $k_i$ and vanishes for $k_i = k_j$, and introduced the constraint explicitly.
The second line follows since $d_{n,{\bf k}}$ (the expression in the parenthesis) is antisymmetric
in the $k_i$, and thus it picks the fully antisymmetric part of 
$z_1^{k_1} \cdots z_N^{k_N}$, reproducing $\psi_{\bf k} ({\bf z})$. The third line is obtained
by shifting summation variables. In doing so, the term $N(N-1)/2$ in the Kronecker $\delta$ is absorbed.

The above holds for arbitrary $n$. We now take the thermodynamic limit $n\gg 1$.
The typical $k_i$ is of order $n$, and thus the exponent in the expression is of order $n$,
and any prefactor polynomial in $n$ is irrelevant, as is the factor $\Delta({\bf z})$. Similarly, the action of 
$\displaystyle \prod_{j>i} (S_i^{-1} - S_j^{-1} )$ produces a subleading factor that can be ignored.
(One way to see this is to note that in the large $n$ limit
the shift operators act as derivatives ($S_i^{-1} - S_j^{-1} \simeq \partial_{k_i} - \partial_{k_j}$)
and bring down subleading terms). Further, we apply to $k_r !$ the Stirling approximation.
Altogether we obtain
\be
Z =  \sum_{\bf k} \delta_{k_1 + \cdots +k_N ,n} e^{-\beta\, {\rm F}({\bf k})+{\cal O} (n^0)}\ ,
\label{Zeta}\ee
where the free energy of the system is, up to a trivial overall constant,
\be
\label{ofho1}
{\rm F}({\bf k})= -T n\ln n + \sum_{i=1}^N \Big( T  k_i \ln k_i - {c \over 2n} k_i^2 - B_i k_i\Big)\ .
\ee

In the large-$n$ limit, quantities $k_i$ and $\rm F$ are extensive variables of order $n$. We will now
transition to intensive variables, that is, quantities per atom. To this end, we define rescaled variables
$x_i$ as
\be
\label{constkx}
k_i = n x_i\ ,\quad  i=1,2,\dots , N\ \ .
\ee
satisfying the constraint
\be
\sum_{i=1}^N x_i = 1 \ .
\label{constt}\ee
In terms of the $x_i$, the non-extensive term $-T n \ln n$ in the free energy cancels and $\rm F$
becomes properly extensive,
\be
{\rm F} ({\bf x}) = n \sum_{i=1}^N \Big( T x_i \ln x_i - {N T_0 \over 2} x_i^2 - B_i x_i \Big) 
= n F({\bf x}) 
\label{Fexten}\ , 
\ee
where we have defined
\be
\label{cnt}
c= N T_0\ ,
\ee
introducing a temperature scale $T_0$.
From now on we will work with the intensive quantities $x_i$ (magnetization per atom) and $F$
(free energy per atom) and will omit the qualifier "per atom".

In the large-$n$ limit the sum in \eqn{Zeta} can be obtained by a saddle-point approximation, as the
exponent is of order $n$, by minimizing the free energy $F({\bf x})$ while respecting the constraint 
$\sum_{i=1}^N x_i = 1$. This can be done with a Lagrange multiplier. Adding the term 
$\l (1-\sum_{i=1}^N x_i)$ to \eqn{Fexten} and varying with respect to $x_i$ we obtain
\be
\label{bvara1o}
 \del_i F_{\l} = T \ln{x_i} - {NT_0} x_i   -B_i -\l =0\ , \qq   i=1,2,\dots ,N\ .
\ee
The Lagrange multiplier $\lambda$ can be eliminated by subtracting one of the relations, say for $i=N$,
from the rest (which is equivalent to solving the constraint and expressing one of the $x_i$, say $x_N$,
in terms of the others). We obtain
\be
\label{bvara1}
T \ln{x_i\ov x_N} - {NT_0} (x_i -x_N) -(B_i -B_N) =0\,  , \quad i=1,2,\dots ,N-1\ ,
\ee
where $x_N$ is determined from the constraint \eqn{constt}.
Also, from \eqn{bvara1o} we obtain the second derivatives
\be
\label{bvara2o}
\del_i \del_j F_{\l} = \Big({T  \ov x_i} - {NT_0}\Big) \d_{ij} \ ,\quad  i,j=1,2,\dots ,N\ ,
\ee
subject to \eqn{constt}.
The above Hessian will be needed later in order to investigate the stability of the solutions. The
simpler form of the equations (\ref{bvara1o}) involving $\lambda$
will also be useful in determining the nature of solutions and in the stability analysis.

\section{Phase transitions with vanishing magnetic fields}

We now put $B_i = 0$ (setting all of the $B_i$ equal is equivalent, as this would be a $U(1)$ field
and would contribute a trivial constant to the energy) and examine the phase structure of the system.
We can collectively write equations \eqn{bvara1o} for $B_i =0$ as
\be
\label{simusu}
T \ln x - NT_0 x =   \lambda\ ,
\ee
dropping the index $i$ in $x_i$ to emphasize that it is the same equation for all $x_i$'s, unlike the case
with generic non-vanishing magnetic fields.
The value of $\l$ is fixed by the summation condition \eqn{constt}.

\begin{figure} [th!] 
\begin{center}
\includegraphics[height= 4.5 cm, angle=0]{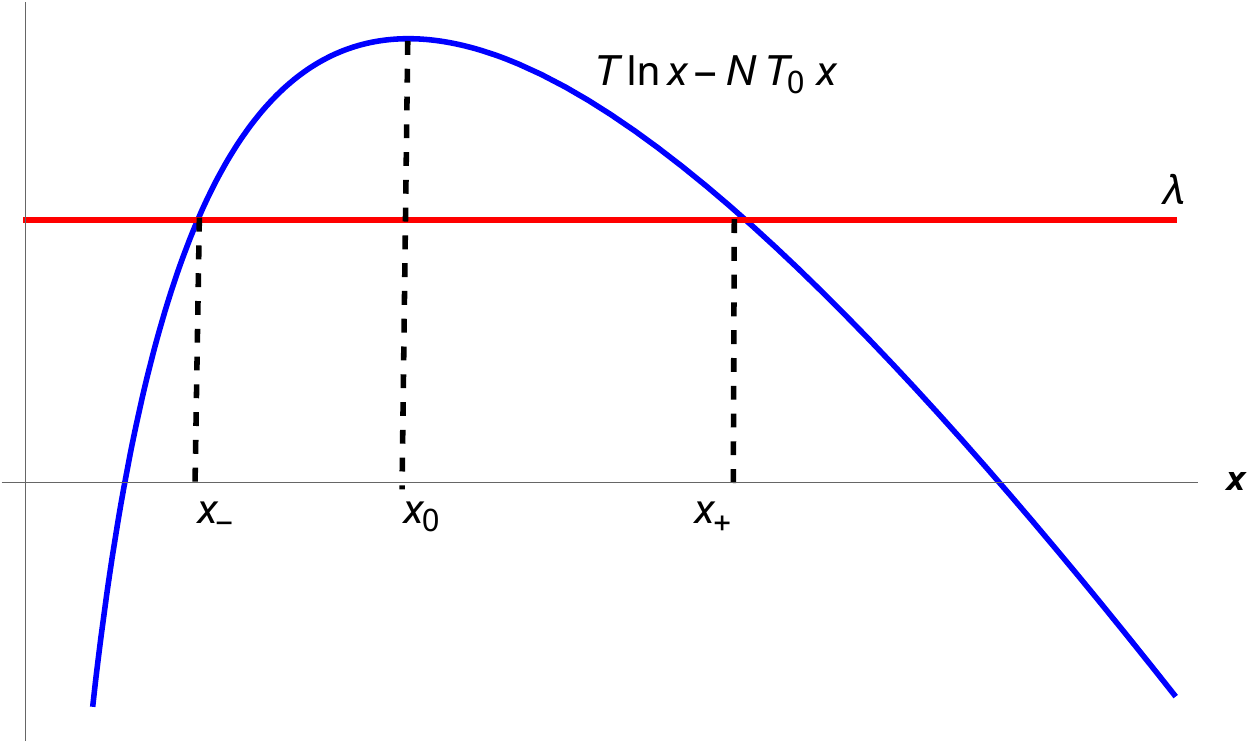}
\end{center}
\vskip -.5 cm
\caption{\small{Plot of the LHS of \eqn{simusu}, with its maximum occurring at $\displaystyle x_0={T\ov NT_0}$.
The intersection with some constant value of the Lagrange multiplier $\l$
occurs at $x=x_\pm$, with $x_-<x_0<x_+$. }}
\label{quasimodo1}
\end{figure}

We note that (\ref{simusu}) always admits the trivial solution $x_i = 1/N$ (for an appropriate $\lambda$),
corresponding to the singlet irrep and an unbroken $SU(N)$ phase. Generically, however,
the above equation has two solutions (see fig. \ref{quasimodo1}). So, each $x_i$ can have
one of two fixed values, $x_-$ or $x_+>x_-$. This means that the dominant irreps are those with $M$
equal rows, where $M$ is the number of $x_i$ having the large stvalue $x_+$ in the solution,
and gives $SU(M) \times SU(N-M) \times U(1)$ as the possible {\it a priori} spontaneous breaking
of $SU(N)$, the subgroup that preserves a matrix with $M$ equal and $N-M$ different and equal diagonal
entries. We will see, however, that stability of the configuration requires that at most one $x_i$ value in
the full solution be $x=x_+$; that is, either $M=0$, corresponding to the singlet, or $M=1$,
corresponding to a one-row YT, a completely symmetric representation.

\vskip 0 cm

\no
Then equations (\ref{bvara1}) with $B_i =0$ become
\be
T \ln{x_i \over x_N} = N T_0  (x_i -x_N)\ , \quad i = 1,2,\dots ,N-1\ ,
\label{xixi}
\ee
where $x_N = 1-x_1 - \cdots -x_{N-1}$ is determined by the constraint in \eqn{constt}.

\no
As argued before from \eqn{simusu}, each $x_i$ can have one of two possible values. Hence,
take $M$ of the $x_i$ to be equal, and the remaining $N-M$ also equal and different. The integer $M$ can
take any value from 0 to $N$, but the values $M=0$ and $M=N$ correspond to the singlet
configuration $x_i = 1/N$ that trivially satisfies (\ref{simusu}).
For $M \neq 0,N$, taking into account the summation to one condition, we set
\be
\begin{split}
& x_i = {1+x\over N} \ ,\qquad  i=1,2,\dots,M \  ,
\\
& x_i = {1-a x \over N}\, ,\quad  i=M+1,\dots, N\ ,\qq a={M\over N-M}\ .
\end{split}
\label{xia}
\ee
Note that, according to \eqn{k1k2}, the $x_i$'s cannot strictly be equal for finite $n$. However, in the
large $n$ limit, differences of ${\cal O}(1/n)$ are ignored.
Further, the choice of the specific $x_i$ that we set to each value is irrelevant, since the saddle point
equations for $B_i =0$
are invariant under permutations of the $x_i$. With the choice \eqn{xia},
$N-M$ of the equations are identically satisfied and the remaining $M$ amount to
\be
T \ln{1+x \over 1-a x} - (1+a ) T_0 x = 0  \ .
\label{xa}
\ee
This transcendental  equation is invariant under the transformation 
\be
x \to -a x\ ,\quad  a\to 1/a\ \ ({\rm equivalently}\ M\to  N-M)\ .
\ee
Thus, without loss of  generality we can choose 
\be
\label{hufhg}
M\leqslant [N/2]\ ,\quad {\rm or}\quad   0< a\leqslant 1\  , \quad x\in (-1,1/a)\ ,
\ee
where $[ \,\cdot\, ]$ denotes  the integer part. 
Solutions with $x>0$ specify an irrep with $M$ equal rows of length, using \eqn{ellk} and \eqn{xia},
\be
\label{xinm}
\ell_i=  {x\ov N-M}\, n + {\cal O}(1)\, , \quad i=1,\dots, M\ ,\qq \ell_i = {\cal O}(1)\, , \quad i=M+1,\dots, N-1\ .
\ee
Instead, an $x<0$ specifies an irrep with $N-M$ equal rows (corresponding to the conjugate representation), with
length given by \eqn{xinm} but with $x$ replaced by
$ x\to -x$.
 The reason is that in this case the $x_i$'s
in the second line of \eqn{xia} are larger than those of the first line and  therefore the roles of
$M$ and $N-M$ are reversed.

\begin{figure} [th!] 
\begin{center}
\includegraphics[height= 4 cm, angle=0]{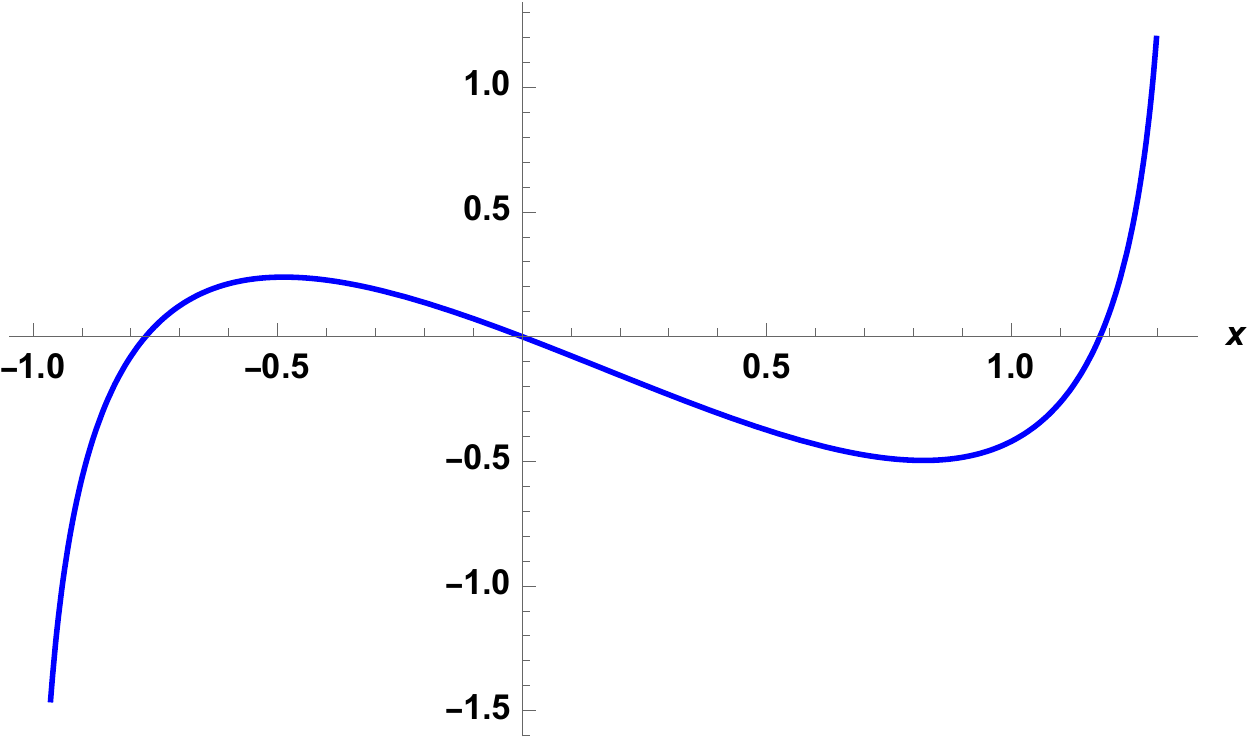}\hskip .5 cm  \includegraphics[height= 4 cm, angle=0]{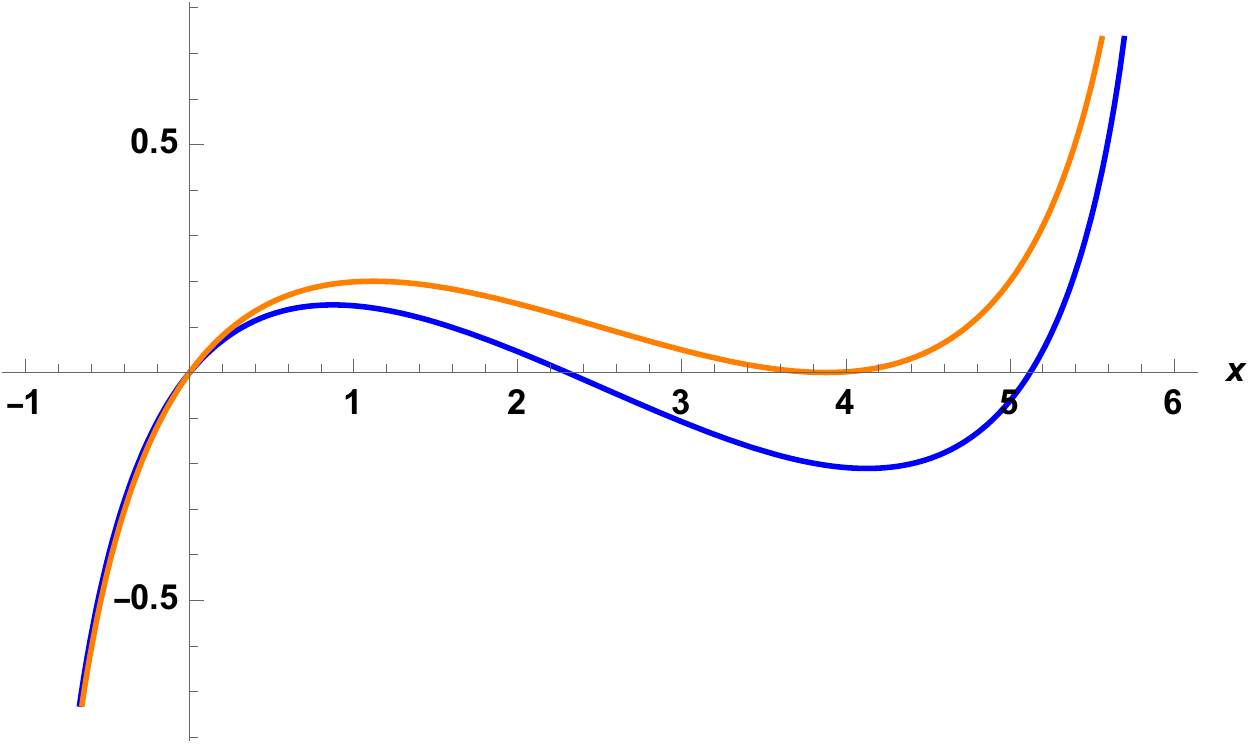}
\end{center}
\vskip -.5 cm
\caption{\small{Plots of the left hand side of \eqn{xa} for $N=7$.  \underline{Left:} $M=3$, $T=0.7$. \underline{Right:}  $M=1$, $T=1.6$ (blue) and for   
$T=T_c\simeq 1.72$ (corresponding to $x=x_c\simeq 3.88$) (orange). 
Temperature is in units of $T_0$.}}
\label{solufigs}
\end{figure}

\no
There are generically either one or three solutions to \eqn{xa} depending on $T$ (see fig. \ref{solufigs}). 
If the temperature is higher than a critical temperature $T_c$, then the only solution is that with $x=0$,
that is, the singlet. If $T < T_c$, then there are two additional solutions.
For $T=T_c$ these two solutions coalesce at $x=x_c$, implying that the $x$-derivate of \eqn{xa} is
zero as well at $x_c$. These conditions are summarized as
\be
\label{hgh2}
\begin{split}
&  {T_c \over T_0} \ln{1+x_c \over 1-a x_c} = (1+a ) x_c\ ,
\\
&
(1+x_c)(1-ax_c )= {T_c\ov T_0}  \equiv t\ .
\end{split}
\ee
Solving the first condition for $t=T_c /T_0$ and substituting into the second we obtain a transcendental
equation that determines  $x_c$
\be
\label{1aa}
{(1+a) x_c\ov (1+x_c)(1-a x_c)} = \ln{1+x_c \over 1-ax_c}
\ee
and from that and the first of \eqn{hgh2} the critical temperature $T_c$.
Alternatively, solving
the second equation in \eqn{hgh2} for $x_c$ and substituting into the first
one yields the transcendental equation for $t=T/T_0$
\be
\label{trat}
\ln\bigg({1\ov a}{1+a +\sqrt{(1+a)^2-4 a t}\ov {1+a -\sqrt{(1+a)^2-4 a t}}}\bigg)
-{1+a\ov 2 a t} \Big( 1-a +\sqrt{(1+a)^2-4 a t}\Big) =0\ ,
\ee
which assumes that $a\leqslant 1$. For $a\geqslant 1$ we simply replace $a\to 1/a$.

\no
For $SU(2)$, $a=1$ is the only possibility, and for $a=1$ the solution of \eqn{trat} is $t=1$. Hence, 
the critical temperature is just $T_0$, i.e. the one in \eqn{cnt}. For generic $a$,
it can be checked that the left hand side of \eqn{trat} is monotonically increasing in $t$
and the equation has a unique solution in the range 
$\displaystyle 1<t<{(1+a)^2\ov 4 a}$ for all $a\leqslant 1$.
For $a$ near $1$ we have
\be
\label{trat12}
t = 1 + {1\ov 6} (1-a)^2 +\dots \ ,
\ee
whereas for $a$ near $0$ 
\be
\label{trat1}
{1\ov  t} = a\big(-\ln {a} + \ln (-\ln {a})\big)+\dots
\ee
and therefore $t\to \infty $.
Hence, the solution to \eqn{trat} varies monotonically between these two limiting  cases.
Note that for the case $M=1$, $a=1/(N-1)$ (which is particularly relevant, as we shall see)
\be
\label{hskjfh1}
M=1:\qq T_c\simeq T_0\,  {N\ov  \ln N} \gg T_0  \ ,\quad {\rm as} \quad N\gg 1\ .
\ee
 In conclusion, $T_c > T_0$ for any group $SU(N)$ with $N\geqslant 3$. For $T>T_c$ the only
solution is the trivial one, $x=0$, while for $T<T_c$ there are two additional solutions $x_1 , x_2$:
two positive ones $0<x_1 <x_c <x_2$ for $T_0 <T< T_c$, and one positive and one negative one 
$x_1 <0<x_2$ for $T<T_0$.
A few generic cases are depicted in Fig. \ref{quasimodo1} where the left hand side of \eqn{xa} is plotted.
\vskip 0 cm

\subsection{Stability analysis and critical temperatures}\label{Staco}

The solutions of \eqn{xa} may be local extrema or saddle points of the action.
To determine their stability we examine the second variation of the free energy.
From \eqn{bvara2o}, $\d^2 F$ is
\be
\label{2var}
\d^2 F = {1\ov 2} \sum_{i=1}^N C_i^{-1}  \d x_i^2 \ ,\qquad C_i^{-1} =  {T\ov x_i}-N {T_0}\ ,\qquad \sum_{i=1}^N \d x_i =0\ .
\ee
Note that this remains the same even in the presence of magnetic fields, so the stability argument below
is fully general.

If all coefficients $C_i$ are positive at a stationary point, then clearly the solution is stable.
If two or more $C_i$'s are negative, on the other hand, it is unstable. Indeed, we can take, e.g.,
$\delta x_{i_1} +\delta x_{i_2}=0$ for two of the negative coefficients $C_{i_1}$ and $C_{i_2}$, and
set the rest of the $\delta x_i$ to zero in order to satisfy the  constraint. Then, an obvious instability arises.
However, if only {\it one} $C_i$ is negative and the rest of them positive, then the solution could still be
stable due to the presence of the constraint. A standard analysis shows that the condition
for stability in this case is\footnote{A nice way to derive this is to view $\d x_i$ as covariant
coordinates on a space with metric $g_{ij} = C_i \delta_{ij}$ of Minkowski signature $(-,+,\dots, +)$.
Then $\d^2 F$ and the constraint become
\be
\d^2 F = g^{ij} \d x_i \d x_j ~,~~~ u^i \d x_i =0 ~~~{\rm with}~~~ u^i =1\ .
\nonumber
\ee
For the space spanned by the
restricted $\d x_i$ to be spacelike (with positive definite metric), $u_i$ must be timelike, and
$u^i u_i = g_{ij} u^i u^j <0$ implies \eqn{c12N}.}
\be
\label{c12N}
\sum_{i=1}^N C_i < 0\ .
\ee

\begin{figure} [th!] 
\begin{center}
\includegraphics[height= 4.5 cm, angle=0]{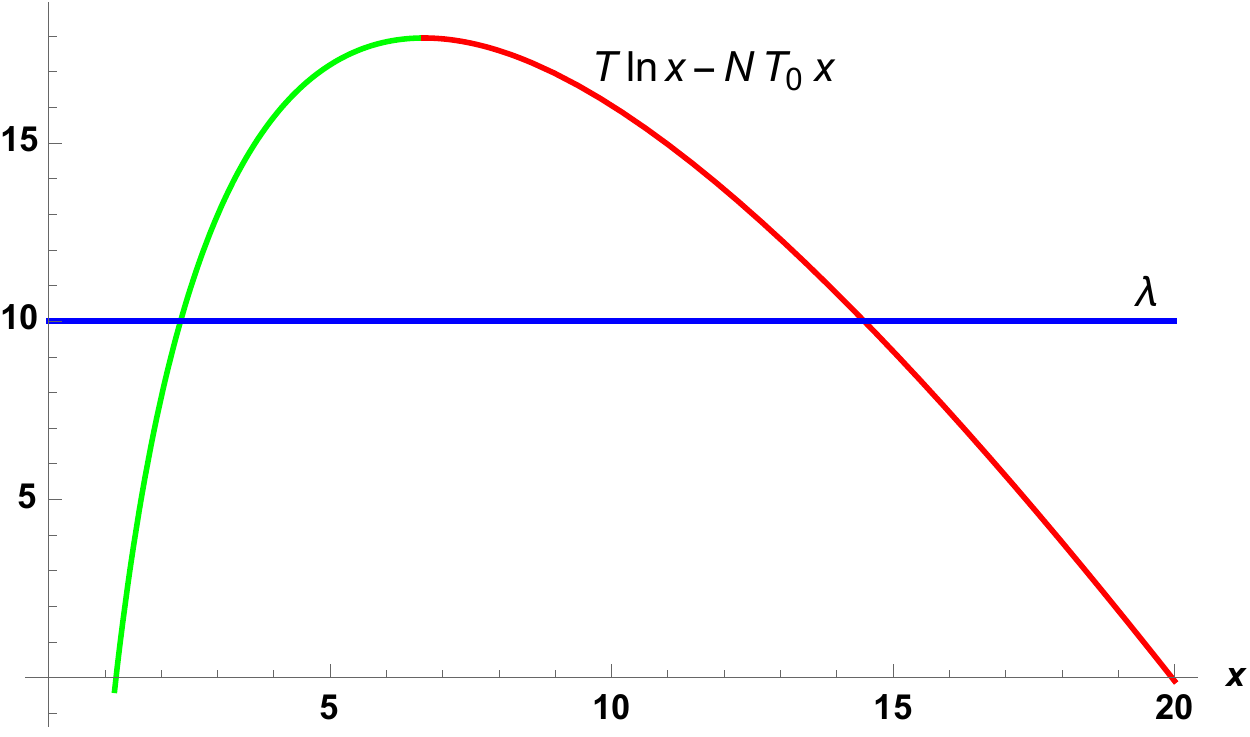}
\end{center}
\vskip -.5 cm
\caption{\small{Plot of \eqn{xixi22},  with its maximum occurring at $\displaystyle x_0={T\ov N T_0}$. The intersection with some constant value of the Lagrange multiplier $\l$ occurs at the values $x_i=x_\pm$, with $x_-<x_0<x_+$. The left (green) part of the curve corresponds to stable solutions, whereas 
the right one (red) to unstable ones.}}
\label{quasimodo2}
\end{figure}

\no
For the case of vanishing magnetic field, $x_i$ satisfy the common equation \eqn{simusu}
\be
T \ln{x} - N {T_0}\,  x = \lambda \ .
\label{xixi22}
\ee
The function on the left hand side is plotted in fig. \ref{quasimodo1}, repeated here as 
fig. \ref{quasimodo2}, and has a maximum at 
$x_0 = {T\over N T_0}$ for all $i$. For $x = x_0$, the coefficient $C_i^{-1}$ determining the
perturbative stability at this value
vanishes, while $C_i^{-1} >0$ for $x < x_0$ and $C_i^{-1} <0$ for $ > x_0$. Therefore, the left branch
of the curve represents {\it a priori} stable points and the right branch unstable ones.

\vskip 0 cm

From the above general discussion, we understand that fully stable configurations correspond either to
choosing all $x_i$'s on the stable branch (all $C_i >0$), or $N-1$ of them on the stable branch and one
on the unstable branch (only one $C_i <0$). This last configuration can still be stable if it satisfies the
condition \eqn{c12N}. So the only potentially stable configurations for zero magnetic field are
the singlet ($M=0$), corresponding to a paramagnetic phase, and the fully symmetric single-row irrep ($M=1$),
corresponding to a specific ferromagnetic phase, with order parameter the variable $x$ in \eqn{xia}
determining the length of the single row.

For the remainder of the zero magnetic field discussion we will focus on the nontrivial solution
$M =1$ and set for the constant $a = M/(N-M)$ the corresponding value
\be
\label{aaa}
a={1\ov N-1}\ .
\ee
For this choice, and with $x_i$ as in \eqn{xia}, the coefficients $C_i$ become
\be
\label{cci}
\begin{split}
 C_1^{-1} =N \, A(x)\ ,\qq   
 C_{i}^{-1} =N \, A(-ax)\ ,\quad i =2,3,\dots, N\ ,
\end{split}
\ee
where we defined
\be
\label{aax}
A(x) = {T\ov 1+x} -T_0\ .
\ee
The stability of the no magnetization solution $x=0$ corresponding to the singlet is easy
to find. In that case $A(0)=T-T_0$, so that for $T<T_0$ the solution $x=0$ is a local maximum
and unstable, and for  $T>T_0$ it is a local minimum.

For $M=1$, it is clear from fig. \ref{quasimodo2} that we must have $C_1 <0$, corresponding to the
single value $x_1=(1+x)/N$ on the unstable branch, and the remaining $C_i >0$, so $A(x) <0$ and $A(-ax) >0$.
Also, $1+ x > 1- ax$, so that $x>0$. Condition \eqn{c12N} must also be satisfied,
\be
\label{dhja}
A(x) + a A(-ax) >0 \quad \Longrightarrow \quad (1+x)(1-a x)< {T\ov T_0}\ ,
\ee
where we used $N-1 = a^{-1}$. Using \eqn{xa} to eliminate $T$ this rewrites as
\be
\label{1a}
 {(1+a)x\ov (1+x)(1-a x)} > \ln{1+x \over 1-ax}\ .
\ee
Note that this has the same form as \eqn{1aa} determining the critical $x_c$. It can be seen that
it is satisfied for $x>x_c$ and violated for $x<x_c$.

As analyzed in the previous section, the existence of solutions with $M=1$ requires $T<T_c$.
For such temperatures, equation \eqn{xa} has two solutions, one larger and one smaller than $x_c$.
Only the solution with $x>x_c$ satisfies \eqn{1a}.
Therefore, for temperatures $T<T_ c$ the solution with $x>x_c$ is stable and a local minimum and
the one with $x<x_c$ unstable.
Referring again to Fig. \ref{solufigs}, the solution to the left of $x=x_c$ 
for $T<T_c$ (blue) 
on the right plot is unstable, whereas the one to the right is stable.

We conclude by noting that at low temperatures $T \ll T_0$, we expect the stable configuration of the
system to be the fully polarized one with $x \simeq x_{\rm max} = N-1$, corresponding to the maximal
one-row symmetric representation with $\ell_1\simeq n$ boxes. Indeed, the solution
of equation \eqn{xa} in that case can by well approximated by
\be
\label{largeT}
x\simeq (N-1) \Big(1- N e^{-N T_0/T}\Big)\ ,\qq T\ll N T_0\ ,
\ee
manifesting a nonperturbative behavior in $T$ around $T=0$.

 \subsection{Phase transitions and metastability}

We saw in the previous section that for $T_0<T<T_c$ both the completely symmetric representation
and the singlet are locally stable. The globally stable configuration is determined by comparing the free
energies of the two solutions. 
The free energy per atom was found in \eqn{Fexten}. For zero magnetic fields, it takes the form
\be
F ({\bf x},T) = \sum_{i=1}^N \Big( T x_i \ln x_i - {N T_0 \over 2} x_i^2  \Big)
\ee
and for the single-row zero magnetic field solution the free energy per atom becomes
\be
\begin{split}
\label{onsh}
 F_{\rm sym}(x,T)= & {T\ov 1+a}\big( a(1+x)\ln (1+x) +(1-a x)\ln(1-a x)\big)
 \\
 &
-{a\ov 2} T_0 x^2 - T \ln N\ .
\end{split}
 \ee
Variation of this expression with respect to $x$ leads to \eqn{xa}.\footnote{{Positivity of the
second derivative 
gives the stability condition \eqn{dhja}, but without any restriction on $a$, that is, the number $M$ of
equal rows in the YT. The reason is that this expression only captures stability under variations of the
length of the YT. Taking also into account perturbations into configurations with additional rows recovers
the general stability condition requiring $M=0$ or $M=1$.}}
For the singlet we have
\be
\label{onshsi}
 F_{\rm singlet}(T)=   - T \ln N\ .
\ee
For a specific temperature $T_1$ and magnetization $x_1$ the singlet and symmetric configurations
will have the same free energy. Equating the two expressions and using \eqn{xa} we obtain
\be
\label{hgh22}
\begin{split}
& T_1 \ln{1+x_1 \over 1-a x_1} = (1+a ) {T_0} x_1\ ,
\\
&
T_1 \ln (1+x_1) = {T_0\ov 2} x_1 (2- a x_1)\ ,
\end{split}
\ee
where we used \eqn{xa} to simplify the expression for $F_{\rm sym}$.
This system is solved by\footnote{
In the next section we will present a method for solving it even in the presence of a magnetic field.}
\be
\label{tempt1}
T_1 = {T_0\ov 2} {N(N-2)\ov (N-1) \ln(N-1)}\ ,\qq x_1 = N-2\ .
\ee
In general
\be
T_0 < T_1 < T_c \ .
\ee
except for $N=2$ where we have $T_0=T_1 = T_c$, implying that for the $SU(2)$ case there is a single
critical temperature. For large $N$
\be
\label{hskjfh2}
T_1\simeq T_0\,  {N\ov 2 \ln N} \gg T_0  \ ,\quad {\rm as} \quad N\gg 1\ .
\ee
Recalling the limiting behavior of $T_c$ in \eqn{hskjfh1}, we note that for large $N$, $T_1\simeq T_c/2$.
For $T_0 <T <T_1$, $F_{\rm singlet} > F_{\rm sym}$, while for $T_1<T <T_c$,
$F_{\rm singlet} < F_{\rm sym}$. The situation is summarized in table \ref{table:1}.
\begin{table}[!ht]
\begin{center}
\begin{tabular}{|c|c|c|c|c|} \hline
  irrep & $T<T_0$ & $T_0<T<T_1$ & $T_1<T<T_c$ & $ T_c<T$
  \\ \hline \hline
{\rm Singlet}  &  {\rm unstable}      &  {\rm metastable}   &  {\rm stable }  & {\rm stable}     
\\ \hline
 {\rm Symmetric} & {\rm stable}    &  {\rm stable} &   {\rm metastable}    &  {\rm not a solution}     
 \\ \hline
\end{tabular}
\end{center}
\vskip -.3 cm
\caption{ \small{Phases in various temperature ranges for $N\geqslant 3$ and their stability characterization.}}
\label{table:1}
\end{table}

\no
Hence, at high enough temperature the only solution is the singlet with no magnetization.
At temperature $T_c$ a magnetized state corresponding to the symmetric irrep (one-row) also emerges,
and is metastable until some lower temperature $T_1$. Between these two temperatures the singlet is
the stable solution. Below $T_1$ and down to $T_0$ the roles of stable and unstable solutions are
interchanged. Below $T_0$ the only stable solution is the one-row symmetric representation.
Hence, we have a spontaneous symmetry breaking as
\be
SU(N) \to SU(N-1)\times U(1)\ .
\ee
 
\begin{figure} [th!] 
\begin{center}
\includegraphics[height= 4.5 cm, angle=0]{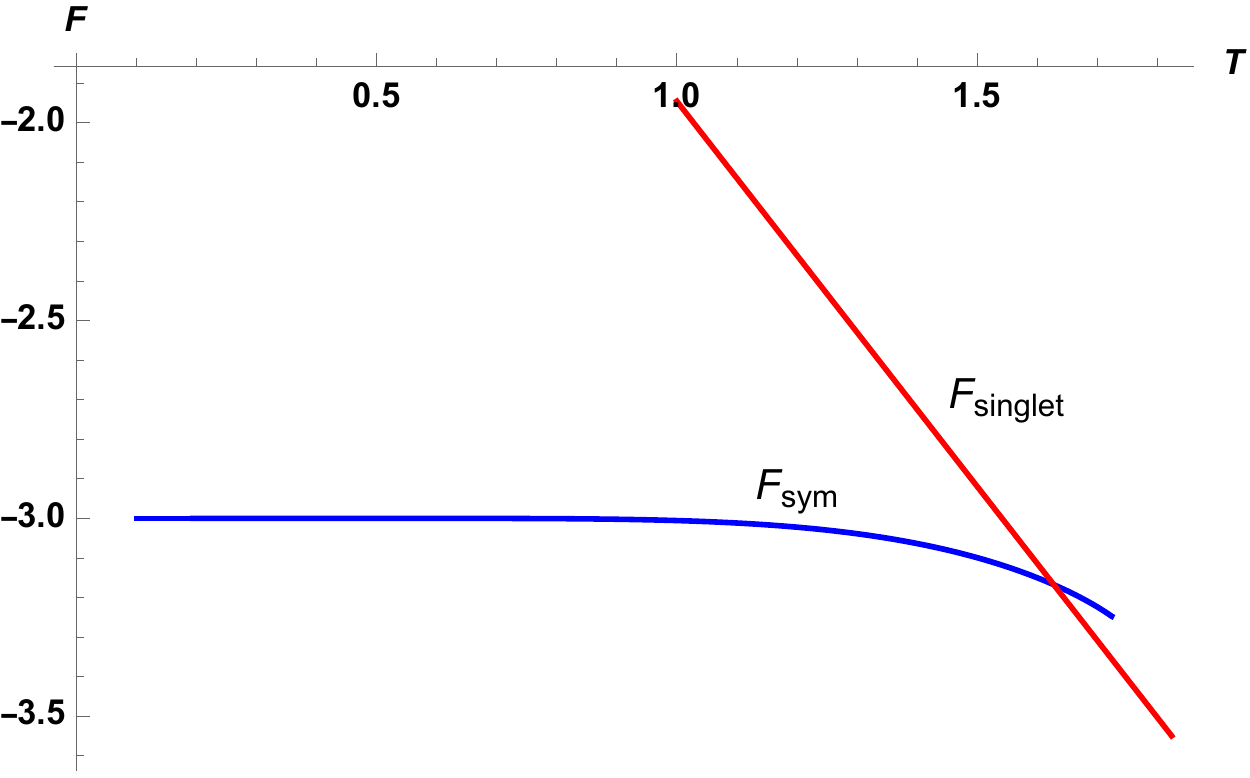}
\end{center}
\vskip -.6 cm
\caption{\small{Plot of the free energy $F$ for $N=7$, for the one-row solution (blue) up to $T_c\simeq 1.72 $ and for the singlet one (red) from $T= 1$,
in units of $T_0$. These cover the temperature range in which we have stability or metastability. 
 The crossover behavior is at $T=T_1\simeq 1.63$, in agreement with table \ref{table:1}}.}
\label{FsymsiNg3}
\end{figure}

\no
Note that the free energy changes discontinuously at $T=T_0$ and at $T=T_c$.
The plot of the free energy is at Fig. \ref{FsymsiNg3}.
The low temperature plateaux in Fig. \ref{FsymsiNg3} for the one-row configuration (blue curve)
is explained by the fact that due to \eqn{largeT} we have
\be
\label{largeT2}
F_{\rm sym} \simeq -{N-1\ov 2} T_0 \Big(1+2 {T\ov T_0} e^{-NT_0/T}\Big) \ ,\qq T\ll N T_0\ .
\ee

\vskip -.4 cm

\no
To better understand the situation, we may use the thermodynamic relations between 
the free energy $F$, the internal energy $U$  and the entropy $S$
\be
F= U-TS\ ,\qq U= -T^2 \del_T\bigg({F\ov T}\bigg)\ ,\qq S= - \del_T F\ ,
\ee
for $F_{\rm sym}$ given by \eqn{onsh} to obtain 
\be
\begin{split}
&
U_{\rm sym}(x) = -{a\ov 2} T_0 \, x^2\ ,
\\
&
S_{\rm sym}(x)=\ln N  -{1\ov 1+a}\big(a(1+x)\ln (1+x) +(1-a x)\ln(1-a x)\big)\ .
\end{split}
\ee
These are readily recognized as the coupling energy and the logarithm of the number of states
per atom
for representations close to the dominant symmetric one (we emphasize that $x=x(T)$ via \eqn{xa}).
For the singlet ($x=0$) we simply have
\be
U_{\rm singlet} = 0\ ,\qq S_{\rm singlet}=\ln N \ ,
\ee
that is, the maximal energy and maximal entropy.

\vskip -.0 cm
\begin{figure} [th!] 
\begin{center}
\includegraphics[height= 4 cm, angle=0]{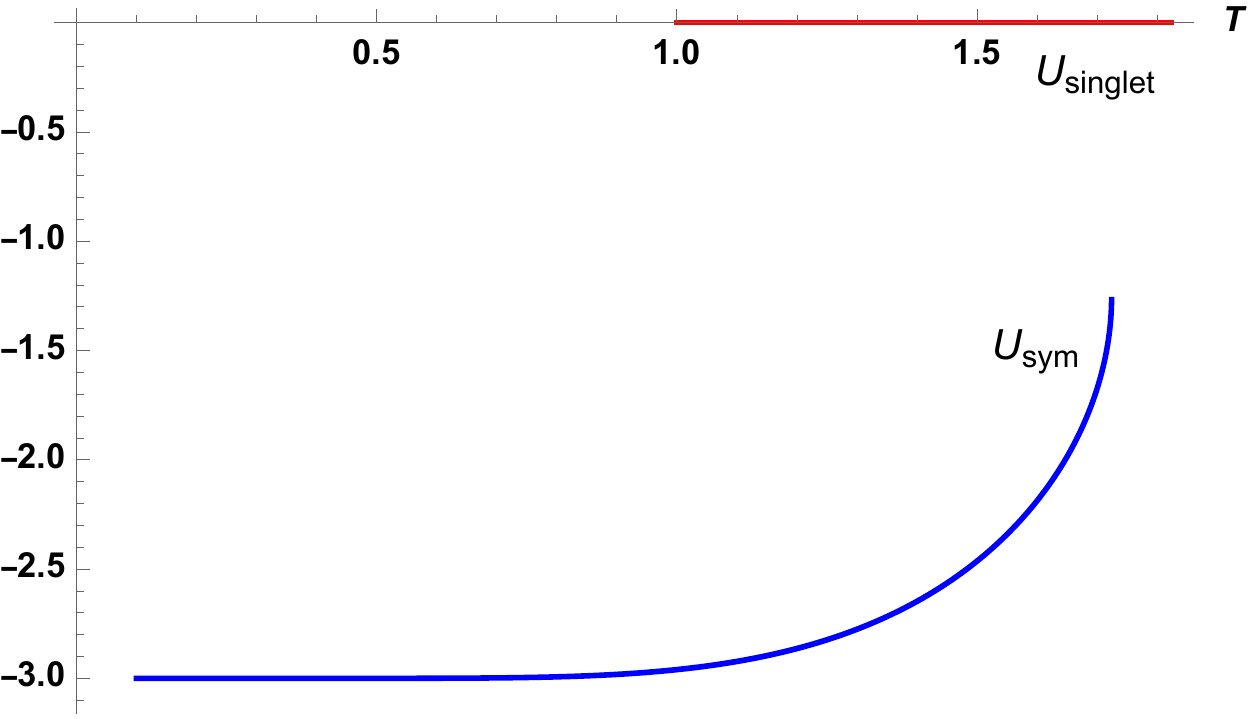}\hskip .8 cm \includegraphics[height= 4 cm, angle=0]{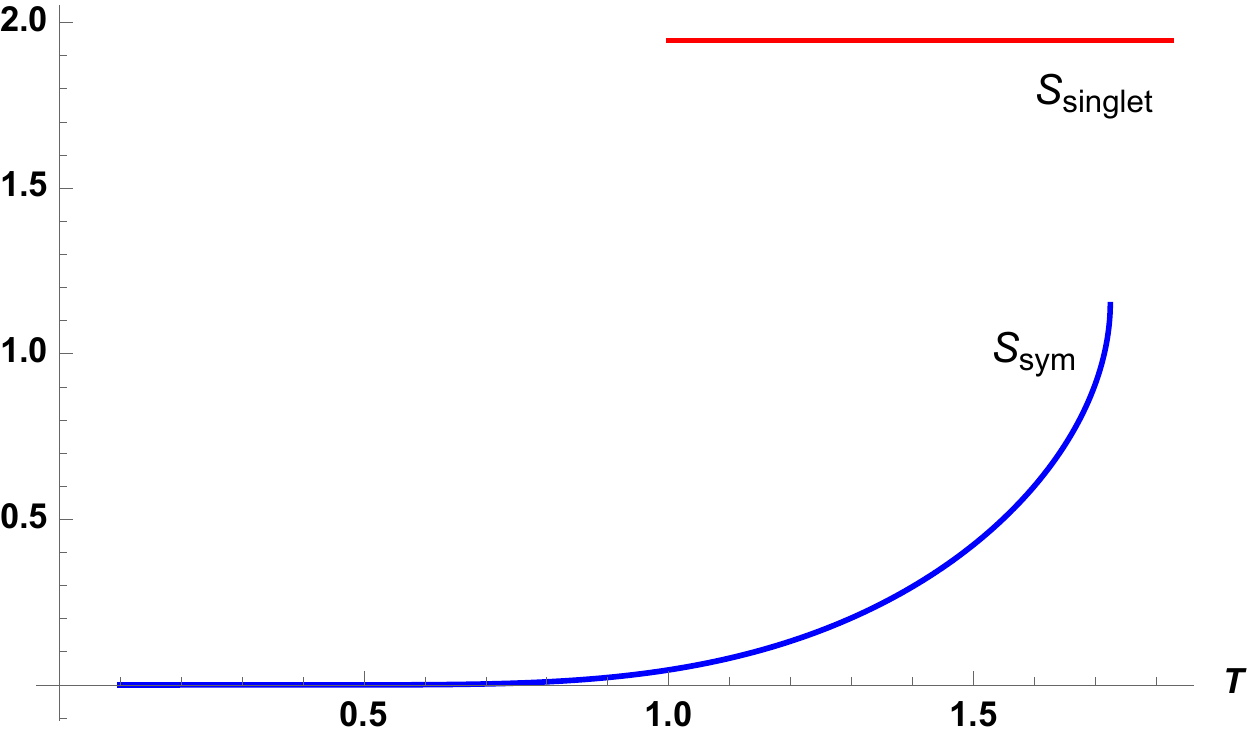}
\end{center}
\vskip -.6 cm
\caption{\small{Plots of $U$ (left) and $S$ (right) for $N=7$.
For the one-row solution (blue) up to $T_c\simeq 1.72 $ and for the singlet one (red) from $T=1$, in units of $T_0$.
The  sharp rise  for $T\to T_c^-$ is according to \eqn{xxcttc}. }}
\label{USNg3}
\end{figure}
\no
Even though the free energy is discontinuous at $T=T_c$, we may still think of the transitions at $T_0$
and $T_c$ as a first order phase transitions, in the sense that the discontinuity of $U$ implies that latent
heat has to be transferred for the phase transition to occur. In detail we have
\be
\begin{split}
&T\to T_c^-:\qq U_{\rm sym}=-{T_0\ov 2(N-1)} x_c^2\ , \quad S_{\rm sym} = \ln N +{x_c\ov 1+x_c} - \ln(1+x_c)\ ,
\\
&T\to T_c^+:\qq U_{\rm singlet}= 0\, , \quad S_{\rm singlet} = \ln N\ ,
\end{split}
\ee
where we simplified $S_{\rm sym}$ by using relations \eqn{hgh2}.
When we transition from below to above $T_c$, we must give energy to the system, which also
increases its entropy (recall that, in the absence of volume effects, $dU = T dS$).
The behavior of the energy and the entropy with temperature is depicted in fig. \ref{USNg3}.

\no
At the intermediate temperature $T=T_1$, given in \eqn{tempt1}, there is the possibility of a first
order phase transition from a metastable to a stable phase. Near that temperature
\be
\begin{split}
& F_{\rm sym}= -T \ln N +{N-2\ov N}\ln(N-1) \big(T- T_1\big) + \dots\ ,
 \\
&  F_{\rm singlet} =  -T \ln N\ .
\end{split}
\ee
This transition is typical in statistical physics where latent heat transfer is involved.
Hence we have that
\be
\begin{split}
&T\to T_1^-:\qq U_{\rm sym}=-\ha {(N-2)^2\ov N-1}\, T_0\ , \quad S_{\rm sym} = \ln N -{N-2\ov N} \ln(N-1)\ ,
\\
&T\to T_1^+:\qq U_{\rm singlet}= 0\, , \quad S_{\rm singlet} = \ln N\ ,
\end{split}
\ee
The transition from metastable to stable configurations near the temperature $T_1$
will not occur spontaneously under ideal conditions, leading to hysteresis.
Only when the system is perturbed, or given an
exponentially large time such that large thermal fluctuations occur, will it transition from a metastable
to a stable configuration. This is reminiscent of the hysteresis in temperature exhibited in certain
materials and in supercooled water \cite{HBAS}.
The discontinuity in $U$ implies that (latent) heat has to be transferred for the phase
transition to occur. When we transit $T_1$ from above the system releases energy, which also lowers
its entropy, the opposite happening when it transit above $T_1$ 
(The term "pseudo phase transition" has been used in the literature for this kind of process).

\no
To understand the nature of the phase transitions that the system can undergo, imagine that we start at a
temperature $T>T_c$ with
the (paramagnetic) singlet state and adiabatically lower the temperature by bringing the system into contact
with a cooling agent (reservoir). If the system is not perturbed, it will stay at the paramagnetic state until
$T=T_0$, where this state becomes unstable, the (ferromagnetic) symmetric representation solution takes over,
and the system undergoes a phase transition releasing latent heat into the reservoir.
Similarly, starting at a temperature below $T_0$ with the (ferromagnetic) symmetric irrep state and raising
adiabatically the temperature, the system will remain in this state if it is not perturbed and will transition to
the singlet at $T=T_c$, absorbing latent heat from the reservoir and undergoing
a phase transition. Clearly the system presents hysteresis, and no unique Curie temperature exists,
since in the range $T_0<T<T_c$ the two phases coexist.

The situation is somewhat different if the system is
isolated (decoupled from the reservoir) while it is in a metastable state (that is, a ferromagnetic state at
$T_1 <T <T_c$ or a paramagnetic state at $T_0 <T <T_1$). A transition to the stable state would involve
exchange of latent heat, which can only be provided by, or absorbed into, the stable phase. However, the
heat capacity of the paramagnetic phase is zero (see figure \ref{USNg3}), so no such exchange can take place
and the system is "trapped" in the unstable phase. This is unlike, say, supercooled water,
where perturbations nucleate the formation of a stable solid state, releasing latent heat into the unstable liquid
phase and raising the temperature until the liquid phase is eliminated or the temperature reaches the point
at which the two phases become equally stable. This feature of our system is somewhat unrealistic, since
we have ignored all other degrees of freedom except $SU(N)$ spins. A realistic system would also have
vibrational degrees of freedom of the atoms, which would serve as a reservoir absorbing or receiving latent
heat and thus enabling transitions from metastable to stable states.

\begin{figure} [th!] 
\begin{center}
\includegraphics[height= 4.5 cm, angle=0]{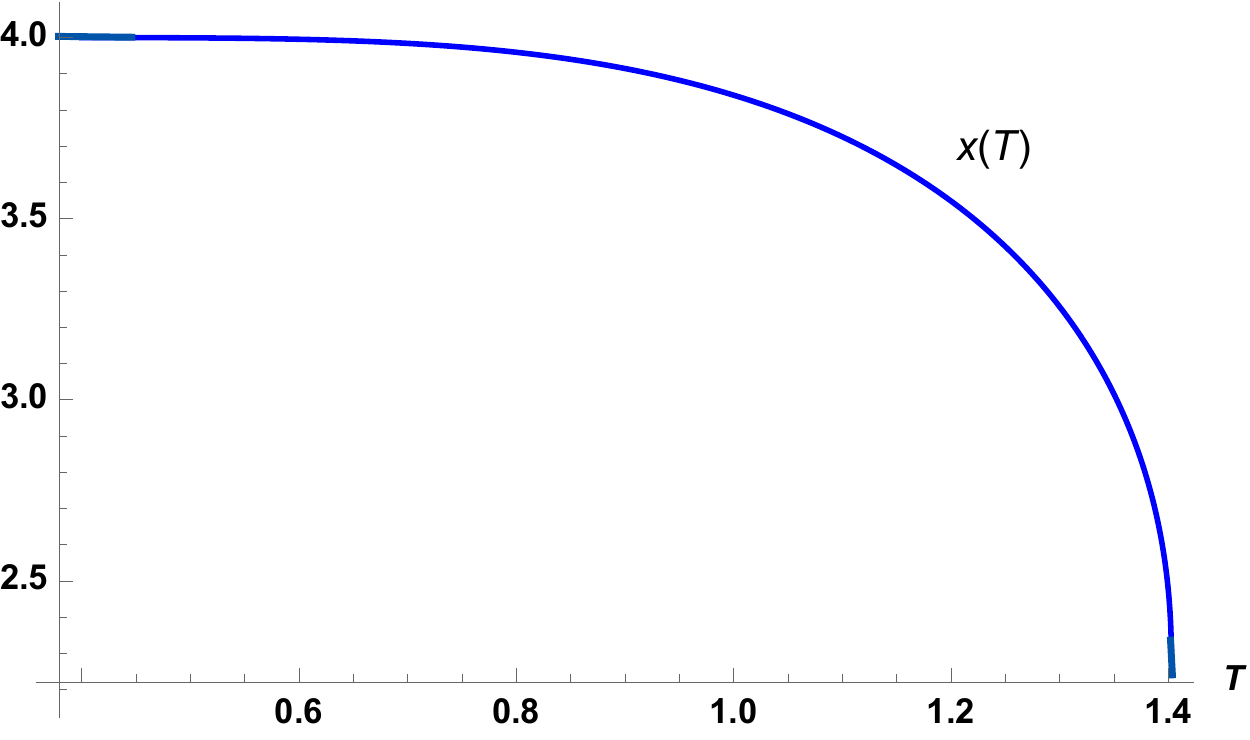}
\end{center}
\vskip -.6 cm
\caption{\small{
Typical plot of $x(T)$ (here for $N=5$) for the symmetric irrep for $0<T<T_c\simeq 1.40 $. For $T=0$
it reaches the maximum 
value $x(0)=N-1$ and for $T\to T_c^-$ it goes sharply to $x_c$ according to \eqn{xxcttc}. }}
\label{magnefig}
\end{figure}

The parameter $x$ is a measure of the spontaneous magnetization and constitutes the order parameter for the phase transition.
Recalling \eqn{xinm}, $x$ is a measure of the length of the single row YT corresponding to the solution.
From \eqn{xa} and \eqn{hgh2} we can deduce that $x(T)$ near $T_c$ behaves 
as\footnote{Positivity of $x_c +1-N/2$
follows from the fact that between the unstable and the stable solutions the left hand side of \eqn{xa}
as a function of $x$ reaches a minimum (see the right plot in fig. \ref{solufigs}.}
\be
\label{xxcttc}
x-x_c  \simeq \sqrt{{(N-1) x_c\ov  x_c +1 -N/2}}\ \sqrt{ {T_c-T\ov T_0}}\ .
\ee
\no
Hence, the deviation from $x_c$ near the critical temperature follows Bloch's law for spontaneous magnetization of materials with an exponent of $1/2$ and an $N$-dependent coefficient. The behavior of $x(T)$ is depicted in fig. \ref{magnefig}.
 
\vskip -0 cm

\paragraph{The $SU(2)$ case:}
The above results are valid for $a\neq 1$ that is for $N\geqslant 3$.  When $a=1$, as in the $SU(2)$ case,
$T_0=T_1=T_c$ and the stable-metastable range in table \ref{table:1} does not exist. 
The transition between the singlet and the symmetric representations occurs at the unique Curie
temperature $T=T_0$ and at $x_c =0$. In this case we have
\ba
\label{txsu2}
& {\rm Symm.\, irrep}\, ( T< T_0){\rm :}\  &x \simeq \sqrt{3} \ \sqrt{T_0 - T\ov T_0} \ ,~~ 
F\simeq -T \ln 2 -{3T_0\ov 4T_0} \big({T_0 -T}\big)^2 \ ,
 \nonumber
\\
& {\rm  Singlet}\, ( T> T_0){\rm :}\ &x=0\ , \quad   F =  -T \ln 2\ ,
\label{jkej}
\ea
which shows that at $T=T_0$ there is a second order phase transition.

\begin{figure} [th!] 
\begin{center}
\includegraphics[height= 4.5 cm, angle=0]{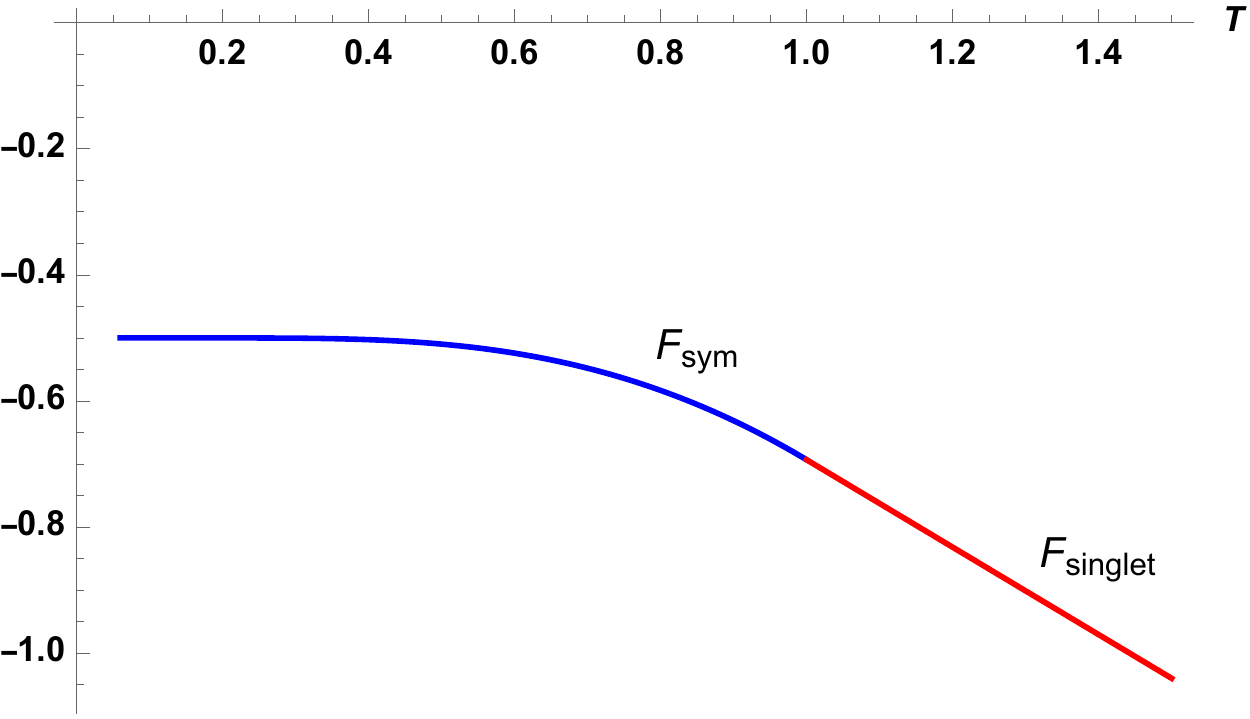}
\end{center}
\vskip -.6 cm
\caption{\small{Plot of the free $F$ for $N=2$. The continuity of the expression and its first derivative between the one-row solution (blue) up to $T=1$ and the singlet one (red) from $T= 1$, in units of $T_0$, is manifest. }}
\label{FsymsiSU(2)}
\end{figure}

\no
Note that the na\"ive limit
$a \to 1$ of the result \eqn{xxcttc} for $a<1$  would not recover the above behavior. The reason is that
the range of validity of \eqn{xxcttc} is $T_0 < T \lesssim T_c$, and it shrinks to zero as $T_c \to T_0$.
The two first-order phase transition points at $T_0$ and $T_c$ fuse into a single second-order
transition in the limit $a \to 1$. Fig. \ref{FsymsiSU(2)} depicts the free energy against the 
temperature for the $SU(2)$ case.
 
\vskip -0 cm

\begin{figure} [th!] 
\begin{center}
\includegraphics[height= 4 cm, angle=0]{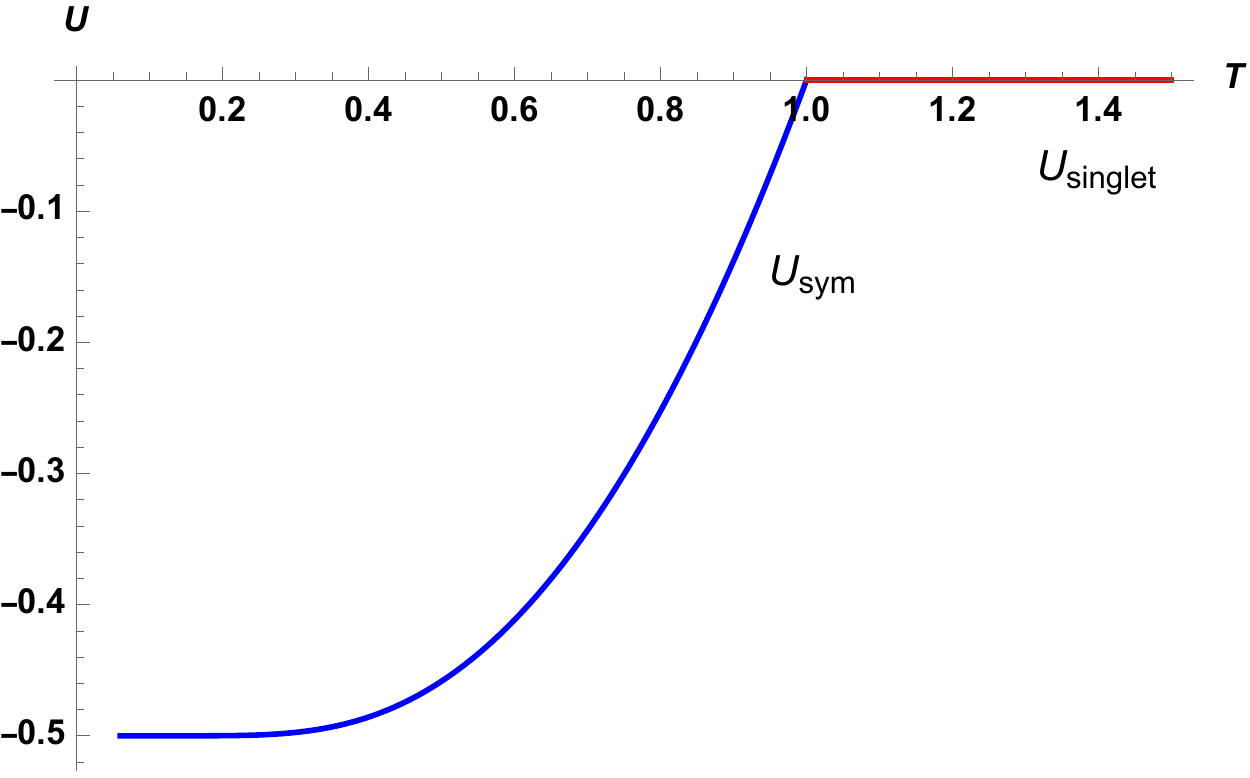}\hskip .8 cm \includegraphics[height= 4 cm, angle=0]{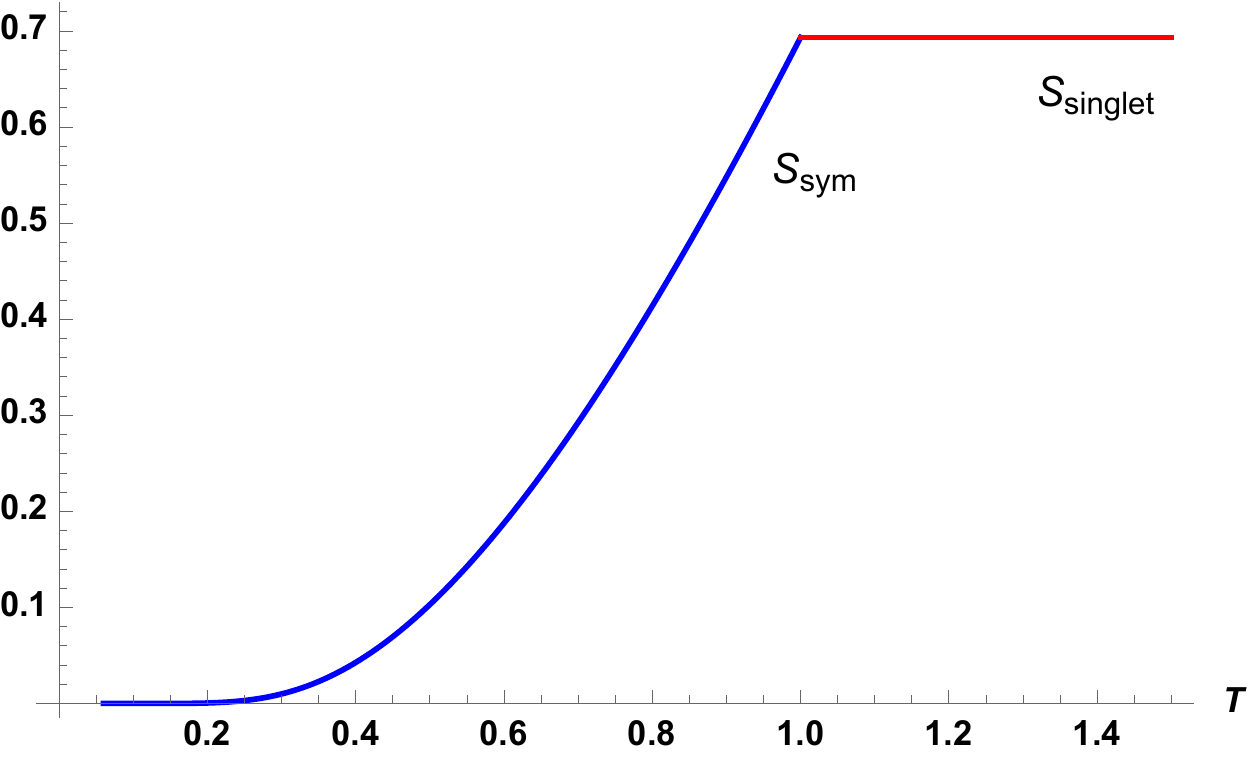}
\end{center}
\vskip -.6 cm
\caption{\small{Plots of $U$ (left) and $S$ (right) for $N=2$ for $T$ in units of $T_0$,
for the magnetized (blue) state up to $T_0$ and the singlet (red) state above $T_0$.}}
\label{USSU(2)}
\end{figure}

\no
The behavior of $x(T)$ near $T=T_0$ follows Bloch's law with an exponent $1/2$, as in the general
$SU(N)$ case, but with a different coefficient.
The internal energy and entropy are continuous but their first derivatives at $T=T_0$ are not,
as depicted in fig. \ref{USSU(2)}.
\vskip -0 cm
For comparison between $SU(2)$ and higher groups, we present in table \ref{table:2} the values
of the critical temperature in units of $T_0$, the critical magnetization as a fraction of the maximal
magnetization $x_{\rm max} =  N-1$ and the coefficient of $x-x_c$ in Bloch's law in \eqn{xxcttc} again as a fraction of
the maximal magnetization, i.e. $K = \sqrt{(N-1)x_c /(x_c +1 -N/2)}/x_{\rm max}$ for the first few values of $N$, as well 
as for $N\gg 1$.
\begin{table}[h]
\begin{center}
\begin{tabular}{|c|c|c|c|c|c|} \hline
  $N$ & $2$ & $3$ & $4$ & $5$  & $N\gg 1$
  \\ \hline
$T_c /T_0$  & $1$     &  $1.0926$   &  $1.2427$  & $1.4027$   &  $N/\ln N$
\\ \hline
 $x_c / x_{\rm max}$  & $0$    &  $0.3772$ &   $0.5071$    &  $0.5749$  &   $1$
 \\ \hline
 K   & $1.7320$  &  $1.2177$  &  $0.9862$ &  $0.8480$  & $\sqrt{2/N}$
 \\ \hline
\end{tabular}
\end{center}
\vskip -.4 cm
\caption{ \small{Behavior near critical point for $N = 2,3,4,5$}, as well as for $N\gg 1$ (the corresponding 
entries are leading).}
\label{table:2}
\end{table}

\section{Turning on magnetic fields}

We now turn to incorporating nonzero magnetic fields in the system. We recall the equilibrium equation
\eqn{bvara1o}
\be
T \ln x_i  - N T_0  x_i  = B_i + \l\ , \qquad i = 1,2,\dots ,N\ ,
\label{xixi2}
\ee
where the $x_i$ sum to $1$ due to the constraint in \eqn{constt}.

\subsection{Small fields}

For small magnetic fields the solution of the coupled equations \eqn{xixi2} can be found as a perturbation
of the solution for vanishing fields. In fact, we can do better than that. We may consider the response of
the system in the presence of generic magnetic fields $B_i$ under small perturbations
$\d B_i$. The state will change as
\be
x_i \to x_i + \d x_i\ ,\quad i=1,2,\dots , N\ ,
\ee
where $\d x_i$ is a perturbation to the solution of \eqn{xixi2}. To linear order we have that
\be
\label{fjgh3}
\left({T\ov x_i} - N T_0 \right) \d x_i = C_i^{-1} \d x_i = \d B_i + \d \l\ .
\ee
Since $\displaystyle \sum_{i=1}^N  \d x_i = 0 $ we obtain the change of $\lambda$ as
\be
\d \l = -{\sum_{i=1}^N C_j \d B_j \ov \sum_{i=1}^N C_j }\ .
\ee
Then, combining with \eqn{fjgh3} we get
\be
\d x_i = C_i \Biggl(\d B_i - {\sum_{j=1}^{N} C_j \d B_j\ov \sum_{j=1}^N C_j } \Biggr)\ ,
\ee
from which the magnetizability matrix $m_{ij}$ obtains as
\be
m_{ij} = {\partial x_i \over \partial B_j}= C_i \,\d_{ij} - {C_i C_j\ov \sum_{k=1}^NC_k}\ ,  
\qq i,j=1,2,\dots, N\ .
\ee
Note that $m_{ij}$ is symmetric and satisfies $\displaystyle \sum_{i=1}^N m_{ij}= 0$ as a
consequence of the fact that the $U(1)$ part decouples.

\no
We may infer the signs of $m_{ij}$ for general $C_i$'s from the stability condition on the configuration. 
Recall that, for stability, either all $C_i$ are positive (including infinity), or only one of them is negative, say $C_1$, and the
rest positive ($x_1$ is the largest among the $x_i$'s). In the latter case, stability further requires that \eqn{c12N}
be satisfied. Using these we can show that 
\be
\label{msi}
\begin{split}
C_1>0: \qquad & m_{11},\, m_{ii}>0\, , \quad m_{1i}<0\,, \quad m_{ij} <0 \ ,
\\
C_1<0: ~\qquad & m_{11},\, m_{ii}>0\, , \quad m_{1i}<0\,, \quad m_{ij} >0   \ ,
\\
C_1^{-1}= 0 :  \qquad &m_{11},\, m_{ii}>0\, , \quad m_{1i}<0\,, \quad m_{ij} =0 \ ,
\end{split}
\ee
where $ i,j=2,3,\dots, N$ and $i\neq j$.
According to \eqn{2var}, the last case above happens for $T = N T_0 x_1$, which is possible only for $T< NT_0$, and
corresponds to one of the solutions of \eqn{xixi2} reaching the top of the function in the LHS of \eqn{simusu},
depicted in fig. \ref{quasimodo1}.

\subsubsection{The singlet}

For the unmagnetized singlet configuration, stable for temperatures $T>T_0$,
$x_i= 1/N$ and $C^{-1}_i=N(T-T_0 )$.
The magnetizability is
\be
\label{hsdkhc}
{\rm singlet}:\quad  m_{ij} = {1\ov N(T-T_0)}\bigg(\d_{ij} - {1\ov N}\biggr)\ .
\ee
As expected, it diverges as $(T-T_0)^{-1}$ 
at the critical temperature $T_0$ where the
configuration destabilizes,
and the signs of its components are in agreement with \eqn{msi}. This determines the linear
response of the system to small magnetic fields, of typical magnitude 
$B$ such that $B\ll T-T_0$.

\subsubsection{The symmetric representation}

For the spontaneously magnetized configuration corresponding to
the symmetric representation $M=1$, stable for $T<T_c$, the $x_i$ are as in \eqn{xia}, with $x$
the solution of \eqn{xa} that corresponds to a stable configuration. 
The components of the magnetization matrix take the form
\be
\begin{split}
& m_{11} = {N-1\ov N \D (x)} \ ,
\\
&
m_{1i}= -{1\ov N \D(x) }\ ,\quad i\neq 1 ,
\\
&
 m_{ij} = {1\ov N A(-ax)} \bigg(\d_{ij} - {A(x)\ov\D(x) }\bigg)\ , \quad i,j\neq 1\ ,
\end{split}
\ee
where $a=1/(N-1)$ and $A(x) = T/(1+x) -T_0$ as before, and
\be
\label{ddd}
\D (x) =   A(-ax) + {1\over a} A(x) = N \left[{T\ov (1+x)(1-a x)} - T_0 \right]\, .
 \ee
As $T\to T_c^-$, \eqn{hgh2} shows that the magnetization diverges.
To compute its asymptotic behavior at $T \simeq T_c$ we use \eqn{xxcttc} to obtain
\be
\label{ddd1}
(1+x)(1-a x)= {T_c\ov T_0} -  \sqrt{ 2 x_c(2 a x_c +a -1)}\ \sqrt{ T_c-T\ov T_0} +\dots \ ,
\ee
which implies 
\be
\D = N {T_0\ov T_c} \sqrt{ 2 x_c(2 a x_c +a -1)}\ \sqrt{ T_c-T\ov T_0} +\dots \ .
\ee
Hence the entries of the magnetizability matrix as $T\to T_c^-$  become
\be
\begin{split}
& m_{11} \simeq {N-1\ov N^2 }\,  { Q\ov \sqrt{T_c-T}}\  , 
\\
& m_{1i}\simeq  -{1\ov N^2 }\,{ Q\ov \sqrt{T_c-T}}\  ,
\\
&
 m_{ij}  \simeq {1\ov N^2(N-1) }\, { Q\ov \sqrt{T_c-T}}\  , \quad i,j =2,\dots, N\ ,
\end{split}
\ee
where
\be
\qq Q = {T_c/T_0\ov \sqrt{ 2 x_c(2 a x_c +a -1)T_0}}\ .
\ee
So the magnetizability diverges as $(T_c -T)^{-1/2}$, and the signs of its components are
in agreement with \eqn{msi}.
We have depicted these in fig. \ref{magnetvarious} below.
\vskip -0 cm
\begin{figure} [th!] 
\begin{center}
\includegraphics[height= 4 cm, angle=0]{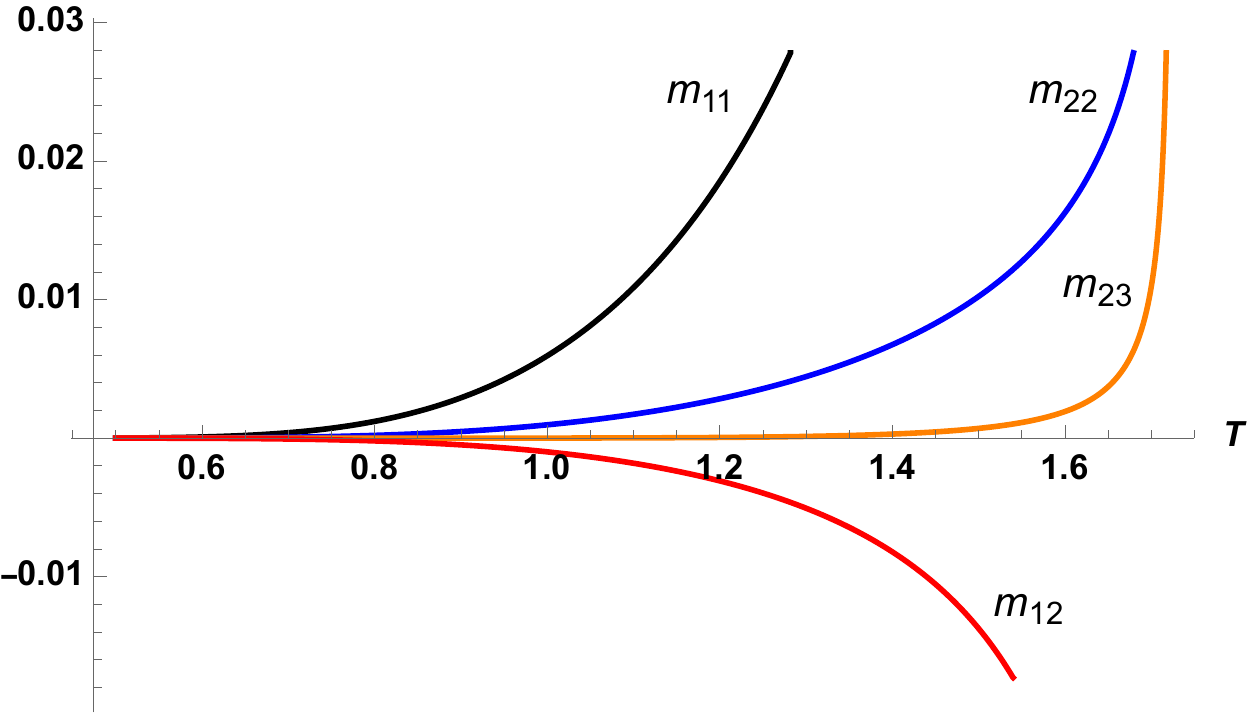}\hskip .3 cm \includegraphics[height= 4 cm, angle=0]{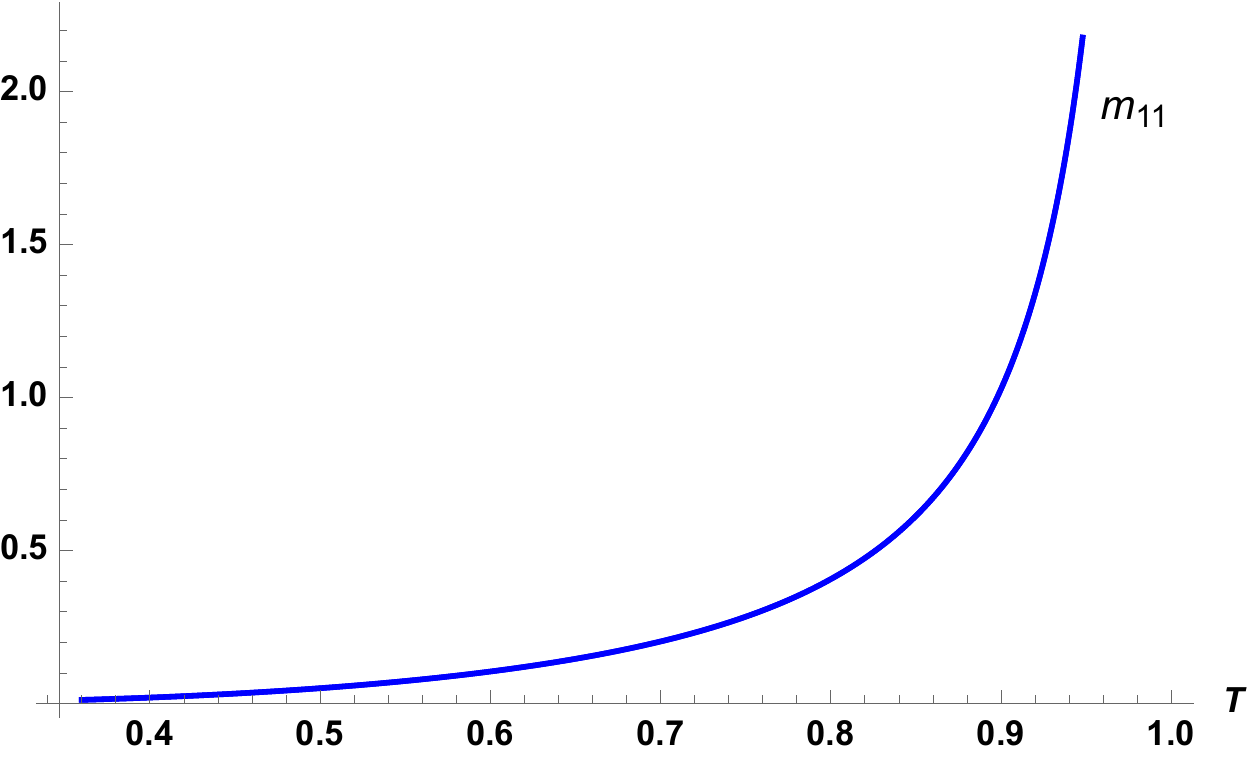}
\end{center}
\vskip -.6 cm
\caption{\small{\underline{Left:} Plots of the four different types of magnetization matrix entries for the symmetric irrep. up to $T=T_c$ for $N\geqslant 3 $.
\underline{Right:}  Same plot for $SU(2)$. }}
\label{magnetvarious}
\end{figure}

\paragraph{The $SU(2)$ case:} This is special since $T_c=T_0$. 
For the singlet solution ($T>T_0$), \eqn{hsdkhc} remains valid for $N=2$, giving
\be
m_{11}=m_{22} =-m_{12}= {1\ov 4} {1\ov T-T_0}\ ,\qq T\to T_0^+\  .
\ee
For the magnetized solution
\be
m_{11}=m_{22}=-m_{12}= {1\ov 4} {1-x^2\ov T-T_0(1-x^2)}\ ,\qq T<T_0\ .
\ee
Using \eqn{jkej} we obtain the magnetizability
\be
m_{11}=m_{22}=-m_{12} \simeq {1\ov 8} {1\ov T_0-T}\ ,\qq T\to T_0^-\ .
\ee
We note that $m_{ij}$ diverges as $|T-T_0 |^{-1}$ on both sides of the critical temperature, unlike
for $N\geqslant2$.
The reason is that the coefficient of $\sqrt{T_c-T_0}$ in the expansion \eqn{ddd1} vanishes for $N=2$
and thus the next order in the expansion, ${\cal O}(T_0-T)$, becomes the leading one. 
We have depicted these in fig. \ref{magnetvarious}.

\subsection{Finite fields}
We now consider the state of the system, given by \eqn{xixi2}, for general non-vanishing
magnetic fields.

The general qualitative picture can be obtained by the same considerations as in the case of
$B_i =0$. For fixed $\lambda$, each $x_i$ satisfies \eqn{xixi2} for a different effective 
Lagrange multiplier $\lambda_i = \lambda + B_i$ and can take two possible values $x_{i\pm}$,
the two solutions of \eqn{xixi2} for fixed $i$. The stability considerations of section \ref{Staco},
which remain valid for arbitrary magnetic fields, determine that at most one of these solutions can lie on
the unstable branch of the curve $x_+$.
So, fully stable configurations correspond either to choosing all $x_{i-}$
on the stable branch, or $N-1$ of them on the stable branch and one on the unstable branch.
The stability condition $\displaystyle \sum_i C_i <0$ in \eqn{c12N} must also be satisfied in the latter case.

\begin{figure} [th!] 
\begin{center}
\includegraphics[height= 4.5 cm, angle=0]{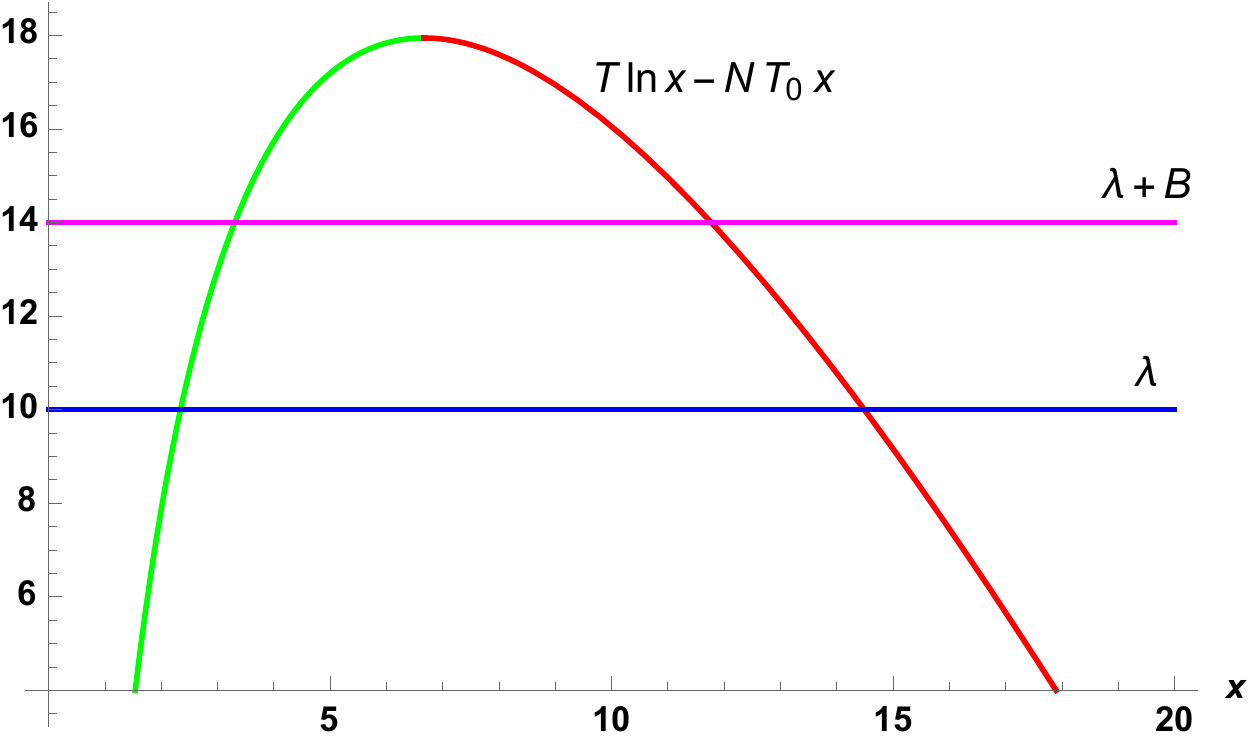}
\end{center}
\vskip -.5 cm
\caption{\small{Plot of \eqn{xixi22},  with its maximum occurring at $\displaystyle x_0={T\ov N T_0}$. We consider a magnetic field $B$ along one direction, 
say for $i=1$.
The intersection with the lines $\l$ and  $\l+B$ occur at two values of $x^i$ on each side of $x_0$ for each line.}
 \label{quasimodo3} 
}
\end{figure}

To gain intuition on the behavior of the system, we focus on the case when only one of the
magnetic fields, say $B_1$, is different from the rest. We can absorb the equal terms 
$B_i$, $i>1$ in the Lagrange multiplier and call $B = B_1 - B_i$. Then the set of equations \eqn{xixi2}
becomes just two distinct equations: one for $x_1$, with RHS $\lambda +B$,
and one for the remaining $N-1$ $x_i$'s, with RHS $\lambda$, as depicted in fig. \ref{quasimodo3}.
\vskip 0 cm
\no
Referring to fig. \ref{quasimodo3}, denote the four solutions of \eqn{xixi2} by $x_\pm$
(at the intersections of the curve with $\l$), and by $x^B_\pm$ (at the intersections with $\l+B$).
For the choice of $B>0$ in the figure we have the ordering $x_-<x_-^B<x_0<x_+^B<x_+$. For $B<0$ the
ordering would change to $x_-^B<x_-<x_0<x_+<x_+^B$. Since we have a magnetic field only in direction
$1$, $x_1$ can take values $x_\pm^B$ while each $x_i$ ($i>1$) can take values $x_\pm$.

For a stable configuration we may choose at most one value at $x_+$ or $x_+^B$.
Thus, we have the following possible cases:

\begin{itemize} 
\item
a) $N-1$ values $ x_-$ and one value $x_-^B$, corresponding to a one-row YT (if $x_-^B > x_-$) or
its conjugate (if $x_-^B < x_-$). This is the deformation of the singlet for $B=0$.

\item
b) $N-1$ values $ x_-$ and one value $x_+^B$, corresponding, again, to a one-row YT. This is 
the deformation of the one-row solution for $B=0$.

\item
c) $N-2$ values $ x_-$, one value $ x_-^B$, and one value $x_+$, corresponding to a two-row YT.  
This is a deformation of the one-row YT solution at $B=0$ by increasing its depth and breaking
the $SU(N)$ symmetry further. Following a similar analysis, these would correspond either to irreps with
two rows, or to irreps with $N-1$ rows, $N-2$ of which have equal lengths. We will encounter such cases below.
\end{itemize}

\no
Although we will not examine in detail the more general configuration of a magnetic field with
$M$ equal components and the remaining $N-M$ equal and distinct, its qualitative analysis is similar.
Fig. \ref{quasimodo3} remains valid, but now with $M$ values of $x_i$ at $\l+B$ and $N-M$ at
$\lambda$. Implementing the stability criterion we have the following cases (for $B>0$):

\begin{itemize} 
\item
$N-M$ values $ x_-$ and $M$ values $x_{-}^{B_i}$, corresponding to a YT with $M$ rows. 
This is the deformation of the singlet for $B=0$.

\item
$N-M$ values $ x_-$, $M-1$ values $x_{-}^{B_i}$, and one value $x_+^{B_i}$, corresponding to
a YT with $M$ rows. 
This is the deformation of the one-row solution for $B=0$.

\item
$N-M-1$ values $ x_-$, $M$ values $x_{-}^{B_i}$, and one value $x_+$, corresponding to
a YT with $M+1$ rows. 
This is the deformation of the one-row solution for $B=0$ by increasing its depth and breaking the
$SU(N)$ symmetry further. 
\end{itemize}
Overall, equality of magnetic field components results in states with equal rows in their YT.

\subsubsection{One-row and conjugate one-row states}

We proceed to investigate quantitatively the effect of a magnetic field in one direction (say, $1$)
in the case where the system is in a one-row solution in the same direction,
or the corresponding conjugate
representation.
That is, we will examine the cases (a) and (b) above (case (c) will be examined in the next subsection).
Then the  equation for the system becomes\footnote{Note that for $a=1$ (the $SU(2)$ case) and
after setting $x={T\ov T_0} y - {B\ov 2 T_0}$  equation \eqn{gfuiy11} becomes
\be
y = {B\ov 2 T} +{T_0\ov T} \tanh y\ , 
\ee
which is the standard expression in phenomenological investigations of ferromagnetism \cite{Feynpara}.
For $a\neq 1$ (the $SU(N)$ case with $N\geqslant 3$)
this is modified by setting $x={2T  y-B\ov (1+a)T_0}$ and results to the generalization 
\be
y = {B\ov 2 T} +{T_0\ov  T} {1\ov \coth y -{N-2\ov N}} \ .
\ee
}
\be
\label{gfuiy11}
T  \ln{1+x \over 1-a x} = T_0\  (1+a)x  + B\ ,\qq  -1<x<{1\ov a}=N-1\ .
\ee
Solutions to this equation with $x>0$ correspond to a single row YT, whereas solutions with $x<0$ 
to its conjugate, that is, a YT with $N-1$ rows of equal lengths. These are related by observing that \eqn{gfuiy11} is invariant
under $x\to -x/a$, $B\to -B$, and $a\to 1/a$, which maps symmetric irreps to their conjugate.

\no
The stability conditions for the solution are determined by the general discussion in the previous
subsection. As before, according to \eqn{cci} $C_1^{-1} = N A(x)$, with $A(x)$ defined in \eqn{aax}, and thus
$C_i^{-1} = N A(-ax)$ for $i>1$. By its definition, $A(x)$ satisfies
\be
\begin{split}
\label{aaxx}
&x>0:\quad A(x)<A(-a x) \ ,\
\\ &x<0:\quad A(x)>A(-a x) \ .
\end{split} 
\ee
The general stability argument requires $C_i >0$ for $i>1$, that is, $A(-ax)>0$.
For $C_1 <0$ we must also have $\sum_i C_i <0$, or $A(x) + a A(-ax) >0$, as it was shown in \eqn{c12N} and \eqn{dhja}. 
Altogether, combined
with \eqn{aaxx}, the stability conditions that cover all cases are 
\be
\label{jjjs1}
x>0: 	\quad A(x) + a A(-a x) >0 \ ,
\ee
which guarantees  positivity of $A(-ax)$ no matter what the sign of $A(x)$, and
\be
x<0: 	\quad A(-a x) >0 \ . 
\label{jjjs2}
\ee
The condition \eqn{jjjs1} above will be satisfied for
\be
\label{xpm0}
x< x_- ~~{\rm or}~~ x > x_+\  ,\qquad x_\pm = {N-2\pm \sqrt{N^2- 4(N-1) T/T_0} \over 2} \ ,
\ee
while \eqn{jjjs2} will be satisfied for
\be
x_0 < x < 0 \ , \qquad x_0 = (N-1)\bigg(1-{T\ov T_0}\bigg)\ .
\label{fg9}\ee
The existence of $x_\pm$, and the condition that $x_0 > -1$, introduce two more temperatures
\be
T_+  = {N^2\over 4(N-1)}\, T_0 > T_0\ ,\qq  T_- = {N\over N-1}\, T_0 > T_0\ .
\ee
For $T>T_+$, \eqn{jjjs1} is satisfied for all $x>0$, while for $T>T_-$, \eqn{jjjs2} is satisfied
for all $x<0$. Note that both $x_\pm \in (-1,N-1)$, and that $x_+>0$, while $x_->0$ for $T>T_0$
and $x_- <0$ for $T<T_0$. In terms of relative ordering of temperature scales,
\be
\begin{split}
&N=3: \quad  T_+ = {9 T_0\over 8} < T_- = {3 T_0\over 2} \ ,
\\
&N=4: \quad  T_+ =T_- = {4 T_0\over 3}\ ,
\\
&N>4:\quad  T_+ > T_-\ .
\end{split}
\ee

\no
We proceed to the analysis of the states of the system. It is most convenient to keep $x$ and $T$ as
the free variables and consider the magnetic field $B$ as a function of $x$ with $T$ as a parameter. Then
\eqn{gfuiy11} implies
\be
\label{gfuiy21}
B(x)= T \ln{1+x \over 1-a x}  - T_0 (1+a)x \ .
\ee
Note that
\be
\label{dbdx}
{dB\ov dx}= {a(1+a)T_0\ov (1+x)(1-a x)} (x-x_+)(x-x_-) = T_0\big(A(x) + a A(-a x)\big) \ .
\ee
Hence, $dB/dx$ is proportional to the stability condition \eqn{jjjs1} for $x>0$. Therefore, $B(x)$ is an
increasing function of $x$ for $T>T_+$ and $T<T_-$, and a decreasing one for $x_-<x<x_+$, with
$x_-$ and $x_+$ as local maxima and minima. The function $B(x)$ is plotted in figure
\eqn{Bola} for various values of the
temperature. The intersection of these graphs with the horizontal at $B$ determines the solutions
for the configuration of the system. 

\vskip -0 cm
\begin{figure} [th!] 
\begin{center}\hskip -1cm
\includegraphics[height= 3.4 cm, angle=0]{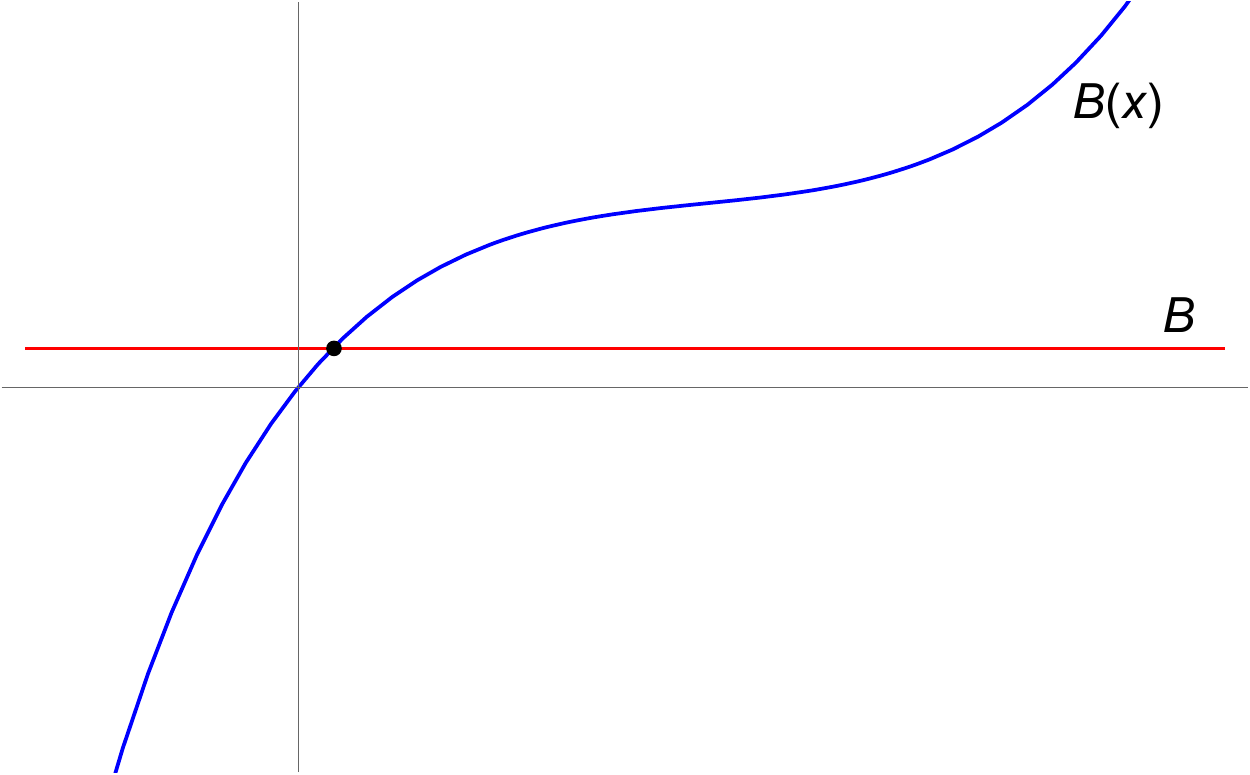} \hskip 0 cm
\includegraphics[height= 3.4 cm, angle=0]{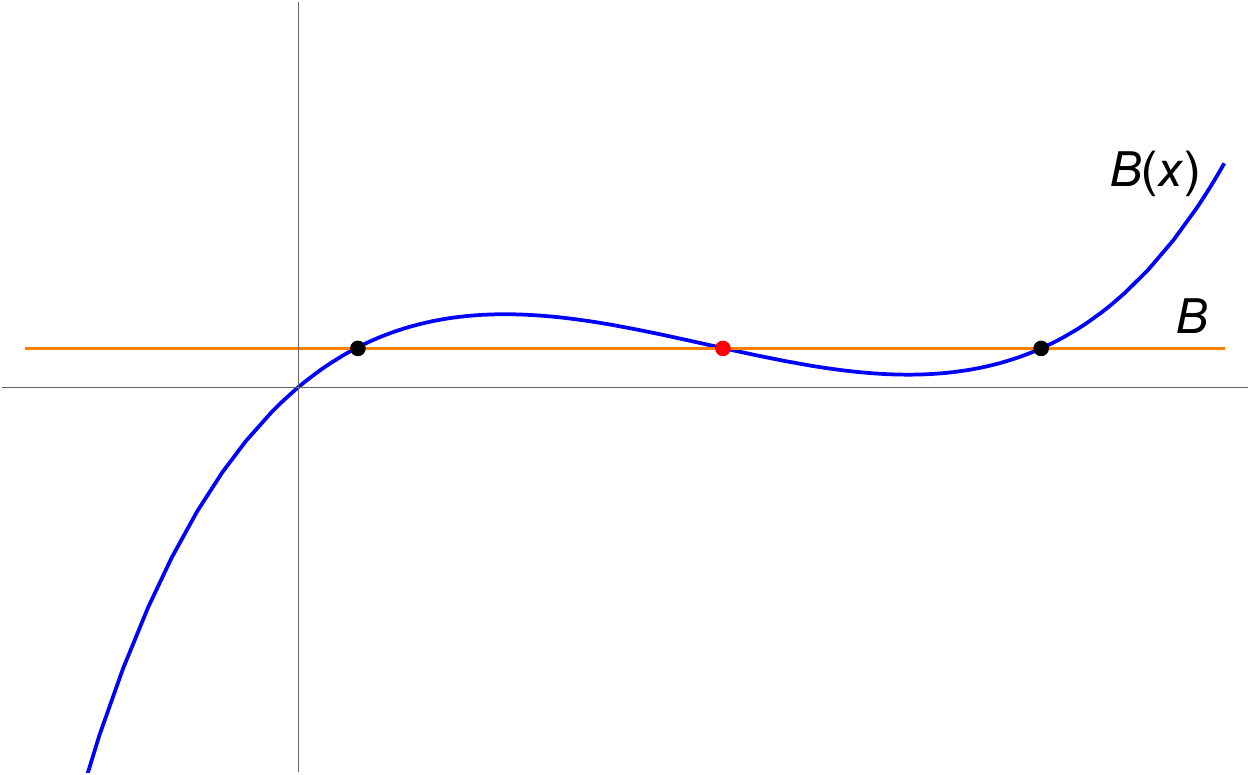}\hskip 0.1 cm
\includegraphics[height= 3.4 cm, angle=0]{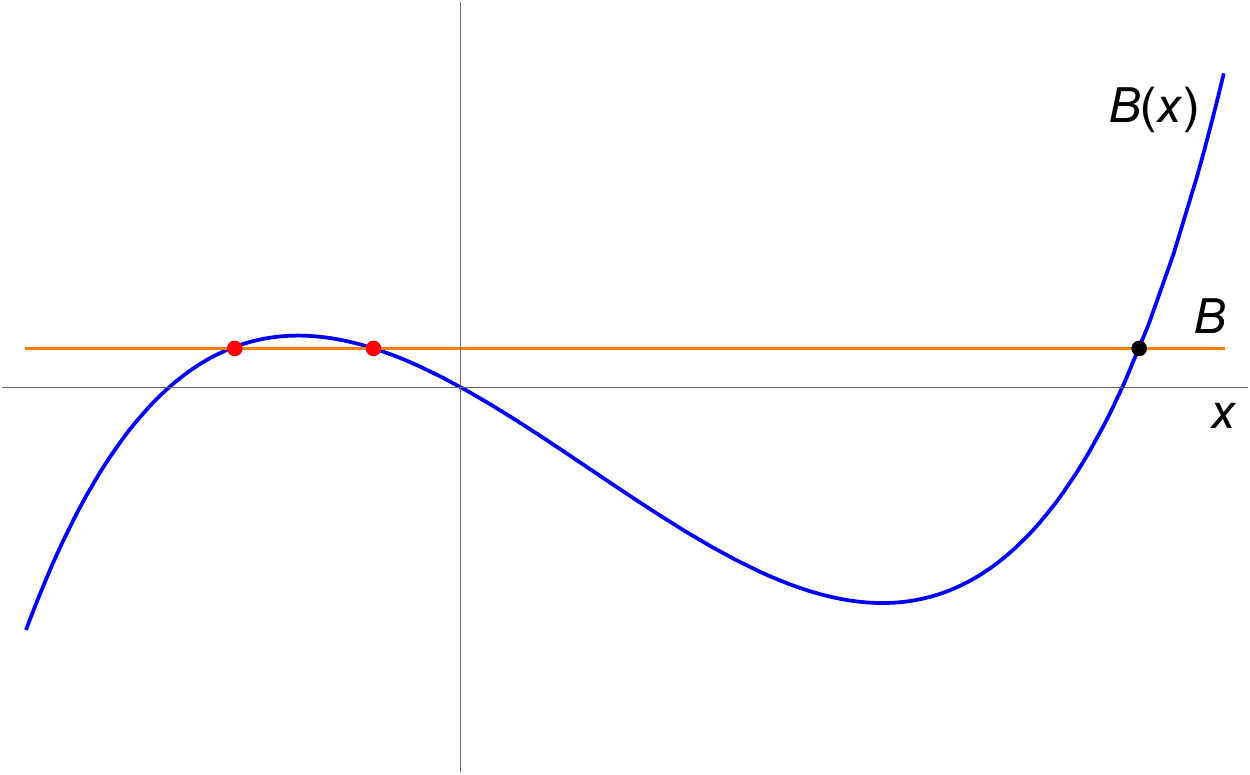}
\end{center}
\vskip -.6 cm
\caption{\small{Plot of  $B(x)$  for  $T > T_+$ (left), $T_0 <T < T_+$ (middle)
and $T < T_0$ (right).}}
\label{Bola}
\end{figure}

The above allow us to determine the stability of solutions for various values of the temperature
and magnetic field. We consider two cases, according to (\ref{jjjs1} and \ref{jjjs2}).

\no
\underline{Positive $x$:} The constraint \eqn{jjjs1} is relevant.
Therefore, when $T>T_+$, all solutions with $x>0$ are stable.
For $T_0<T<T_+$, stability singles out solutions with $0<x<x_-$ and $x> x_+$.
Finally, for $T<T_0$, since then $x_-<0$, stability requires that  $x>x_+$.

\no
\underline{Negative $x$:} The constraint \eqn{jjjs2} is now relevant, or equivalently $x>x_0$.
Since $x<0$, we conclude that no stable solutions exist for $x_0>0$, or $T<T_0$. For $x_0<-1$, or
$T>T_-$, all $x<0$ solutions are stable. Finally, for intermediate temperatures $T_0<T<T_-$, we have stability for $-1<x_0<x<0$.

\no
The above are tabulated in table \ref{table:3} below (we assume that $N>4$ so that $T_+>T_-$):
\begin{table}[!h]
\begin{center}
\begin{tabular}{|c|c|c|c|c|} \hline
  $x$ & $T<T_0$ & $T_0<T<T_-$ & $T_-<T<T_+$ & $ T_+<T$\\
\hline \hline
$x<0$  &  {\rm none}      & $-1<x_0<x$   &  {\rm all}  & {\rm all}     \\ \hline
$x>0$ & $x>x_+$    & $x<x_-$\ \&\ $x>x_+$  &  $x<x_-$\ \&\ $x>x_+$   &  {\rm all}      \\ \hline
\end{tabular}
\end{center}
\vskip -.3 cm
\caption{\small{Stable solutions for various ranges of $T$ and $x$ and for $N>4$. }}
\label{table:3}
\end{table}

\begin{figure} [th!] 
\begin{center}
\includegraphics[height= 9 cm, angle=0]{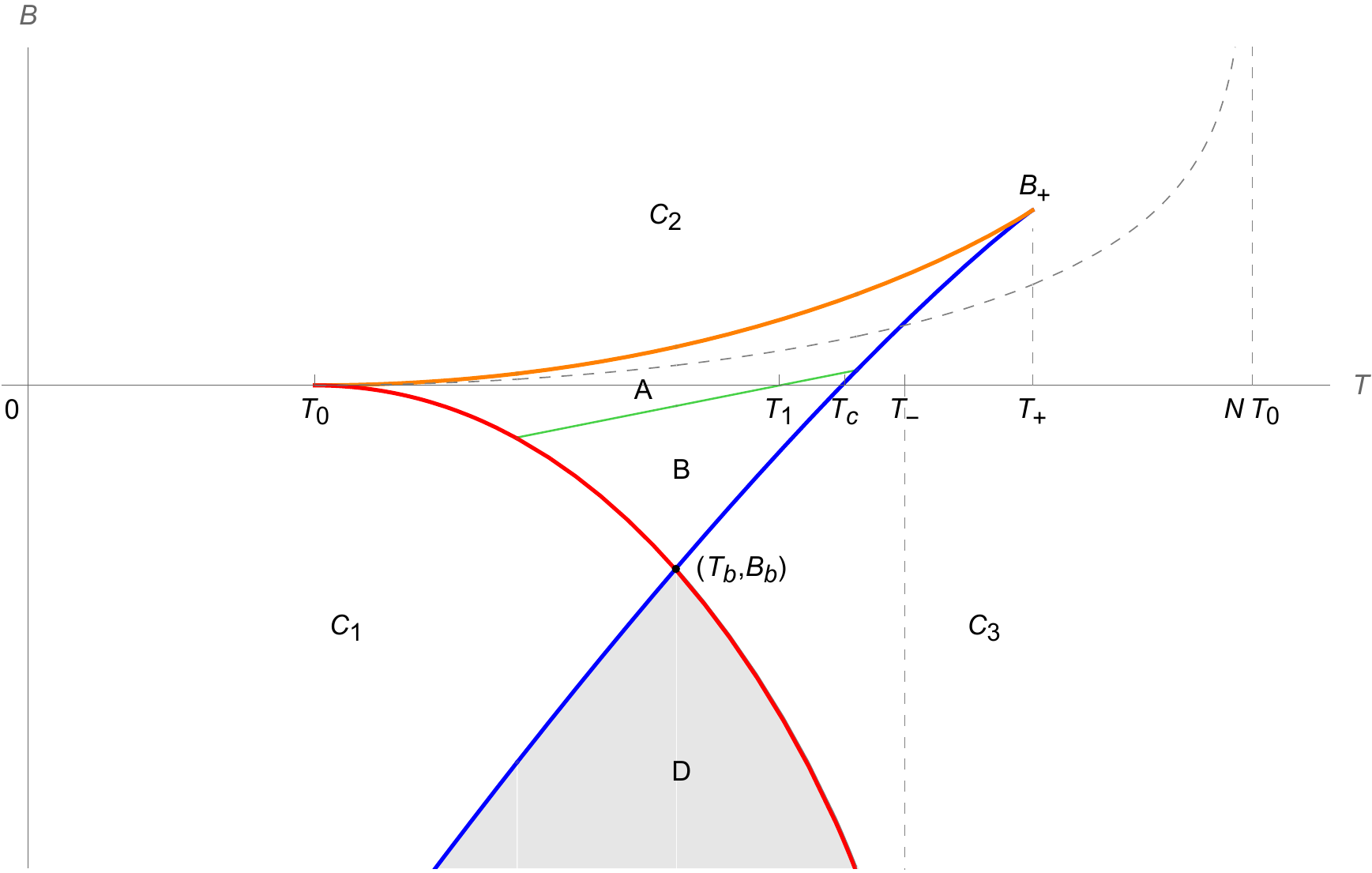}
\end{center}
\vskip -.5 cm
\caption{\small{The phase diagram of the system for generic $N>4$. The (orange) curve separating
regions $A$ and $C_2$ is $B(x_- )$; the (blue) curve separating regions $A,B,C_1$ and $C_3, D$ is
$B(x_+ )$, intersecting the $T$-axis at the $B=0$ critical temperature $T=T_c$;
and the (red) curve separating regions $A,B,C_3$ and
$C_1 , D$ is $B(x_0 )$ and it asymptotes to the vertical $T = T_-$. Crossing any
of these lines precipitates
a discontinuous change in the magnetization, i.e. the order parameter $x$.
The (green) straight line separating regions
$A$ and $B$ is the metastability frontier of the two coexisting phases inside these regions
(see appendix A.1); crossing it exchanges the metastable and the absolutely stable states, and its
intersection with the $T$-axis is the $B=0$ critical temperature $T=T_1$. Regions $C_{1,2,3}$ constitute
one continuous phase with nonzero magnetization (except at $B=0$ and $T>T_c$), all points being
accessible through continuous paths in the $B-T$ space, while regions $A,B$ and $D$ are separated
from $C_{1,2,3}$ by discontinuous transitions in the order parameter. The (gray)
dashed curve from $T_0$ to its vertical asymptote at $T = N T_0$ separates "broken-like" and
"unbroken-like" configurations but otherwise mark no sharp phase transition.
The shaded region $D$ corresponds to a two-row (double magnetization) phase.
For $N=3$ the phase diagram remains qualitatively the same with the order of $T_+$ and $T_-$ interchanged, while for $N=4$, we simply have $T_+=T_-$.}
} \label{TBphase} 
\end{figure}

We can now investigate the existence of stable solutions for the full range of values of the
temperature and the magnetic field. The complete analysis is relegated to appendix A. The results are
summarized in the temperature-magnetic field phase diagram of figure \ref{TBphase}, presented
for a generic value for $N>4$. The phase diagram is qualitatively the same for $N=3$ and $N=4$,
changing only for $N=2$. The only difference
is that, for $N=4$, $T_+ = T_-$, while for $N=3$ $T_+ < T_-$. This does not affect the general features
of the diagram, simply shifting the critical point vertex $(T_+ , B_+ )$ to the left of the bottom asymptote $T=T_-$, for $N=3$, or on top of it, for $N=4$.

Each region in the $T\bb-\bb B$ plane depicted in the figure corresponds to a discrete phase of the system.
Moving within these regions without crossing any of the critical lines interpolates continuously
between configurations.
In the connected regions $C_1$, $C_2$, and $C_3$ there is a unique one-row configuration at each point
$(T,B)$. In regions $A$ and $B$ inside the curvilinear triangle there are {\it two} locally stable
configurations at each point, one absolutely stable and the other metastable, with the line separating $A$
and $B$ being the border of metastability where the two phases have equal free energy.
In region $D$ there are {\it no} stable one-row solutions, signifying that a two-row solution must
exist there. The lines separating regions $C_{1,2,3}$ and the
other regions are phase boundaries, the configuration changing discontinuously as we cross a boundary.

The dashed curve for $T_0<T<N T_0$ represents configurations with $C_1 = \infty$, that is, $A(x)=0$.
This corresponds to points where $x_1 = (1+x)/N$ reaches the top of the curve in fig. \ref{quasimodo3},
transiting from the unstable to the stable branch of the curve or vice versa. For such points,
$x= T/T_0 -1$, and \eqn{gfuiy21} gives $B$ on this curve as
\be
B(T) =  T \ln {(N-1) T \over N T_0 -T} -{N  (T-T_0 )\over N-1}\ .
\ee
Configurations to the left of this curve are in a "broken-like" $SU(N)$ phase, with one of the solutions of \eqn{xixi2}
in the unstable branch of the curve in fig. \ref{quasimodo3}, while those to the right of the curve
are in an "unbroken-like" phase, with all solutions on the stable branch. For $B=0$ these are the true
spontaneously broken or unbroken phases of the system. A nonzero magnetic field breaks
$SU(N)$ explicitly, and the dashed line represents a soft phase boundary, which must be crossed
to transit between the two phases as we move on the $T\bb-\bb B$ plane. The physical signature
of crossing this boundary is that the off-diagonal elements of the magnetizability $m_{ij}$,
with $i\neq j \neq 1$, change sign, vanishing on the boundary (see \eqn{msi}).

\no
The line separating regions $B$ and $C_3$ intersects the $T$ axis at the critical temperature $T_c$
found in the $B=0$ section.
Points $(T_0, 0)$ and $(T_+, B_+)$, with
\be
B_+ = B(x_+ (T_+ )) = {N\over 2(N-1)} \left( {N\over 2} \ln (N-1) -N +2 \right) \ .
\ee
are critical points, while point $(T_b , B_b )$ at the lower tip of region B,
satisfying the transcendental equation
\be
B (x_0 (T_b ) ) = B(x_+ (T_b )) \equiv B_b
\ee
is a multiple critical point, connecting several different configurations: one one-row state in $C_1$,
two in B, and one in $C_3$, as well as a two-row state in D, and possible two-row states in
the other neigboring regions (see next section). One of the states in B, and possibly other
one-row or two-row states, are metastable.

\subsubsection{Two-row states and their ($N-1$)-row conjugates}

As we have seen, for $T < T_-$ and for sufficiently negative magnetic fields there is no stable solution
to \eqn{gfuiy11}, and thus no state corresponding to the one-row YT symmetric representation. From the
general analysis of subsection 4.2, we expect the solution to be the only other allowed configuration,
that is, a state corresponding to a two-row YT. In this subsection we recover this solution and check
its stability. 

\no
We consider a configuration with two (generally unequal) lengths $x_1$ and $x_2$ and an applied
magnetic field in the $x_1$-direction, representing the generic breaking pattern 
\be
SU(N)\to SU(N-2)\times U(1)\times U(1)\ ,
\ee
of the $SU(N)$ symmetry. This includes a spontaneous breaking of $SU(N)$ in addition to the
dynamical breaking $SU(N) \to SU(N-1) \times U(1)$ due to the magnetic field. We note that in the
special case $x_1 = x_2$ the symmetry breaking pattern would be
$SU(N) \to SU(N-2)\times SU(2)\times U(1)$, but as we shall demonstrate this pattern is never realized
in the present case of a magnetic field in a single direction.

\no
The $x_i$ must satisfy the system of equations
\be
\label{xixi2oo}
\begin{split}
&T \ln{x_1 \over x_N} = N {T_0}  (x_1-x_N) + {B}\ ,
\\
&T \ln{x_i \over x_N} = N {T_0}  (x_i-x_N) \ , \quad i = 2,3,\dots ,N-1\ ,
\end{split}
\ee
where $x_N$ is determined from the constraint in \eqn{constkx}. 
We write
\be
\begin{split}
&
x_1 = {1+x\ov N} \ , \qq x_2=  {1+y\ov N}\ ,
\\
&  x_3 =\dots = x_N = {1-\a(x+y)\ov N}\ ,\quad \a= {1\ov N-2}\ . 
\label{xydefs}\end{split}
\ee
For $y=-\a x/(1+\a) = -a x= -x/(N-1)$ this ansatz reduces to the one for the one-row solution.
\eqn{xixi2oo} gives rise to the system of transcendental equations
\be
\label{lnxy1}
\begin{split}
& T \ln{1+x \over 1-\a (x+y)} = {T_0}  \big(\a y + (1+\a)x\big)  + {B}\ ,
\\
& T \ln{1+y\over 1-\a (x+y)} = {T_0}  \big(\a x + (1+\a)y\big)  \ .
\end{split}
\ee
\no
The free energy of the configuration is
\be
\begin{split}
\label{seffxy}
\hskip -1cm F(x,y,T) = &\,{T_0\ov 2 N(N-2)}\Big((x-y)^2 - N(x^2+y^2)\Big) +{T\ov N}\bigg((1+x)\ln(1+x) 
\\
&\hskip -1.5cm +(1+y)\ln(1+y)+(N-2-x-y)\ln\Big(1-{x+y\ov N-2}\Big)\bigg) -{B\ov N}\, x -T\ln N\ .
\end{split}
\ee

The transcendental equations \eqn{lnxy1} will be solved numerically. The full analysis of solutions and
their stability is relegated to the Appendix. Here we simply state the results and present relevant plots.

\no
We consider temperatures $T<T_0$ for which spontaneous magnetization exists. 
As discussed earlier, for such temperatures and negative magnetic fields we expect the state to be in a
two-row stable state, which includes the posssility of "antirows," that is, $N-2$ equal rows plus
an additional row. Further, such states may coexist with a one-row state and be either globally stable
or metastable.

\no
All cases refer to plots in Fig. \ref{2-rowsTB2}. The blue and orange curves represent the solutions of
the first and second equation
in \eqn{lnxy1}, resp. The orange line $y=-ax$, in particular, solves the second equation in the system
\eqn{lnxy1} while the first one reduces to the one for the one-row configuration \eqn{gfuiy11}.
Intersections of blue and orange lines represent the solutions the \eqn{lnxy1}. Only locally stable
solutions are considered.

\begin{itemize}

\item
\underline{$B>0$}: 
We recover the known one-row solution on the $y=-ax$ line for $x>0$. There is also a metastable
two-row solution with $x<0,y>0$.

\vskip -0 cm
\begin{figure} [th!] 
\begin{center}
\hskip -0.2cm\includegraphics[height= 5.2 cm, angle=0]{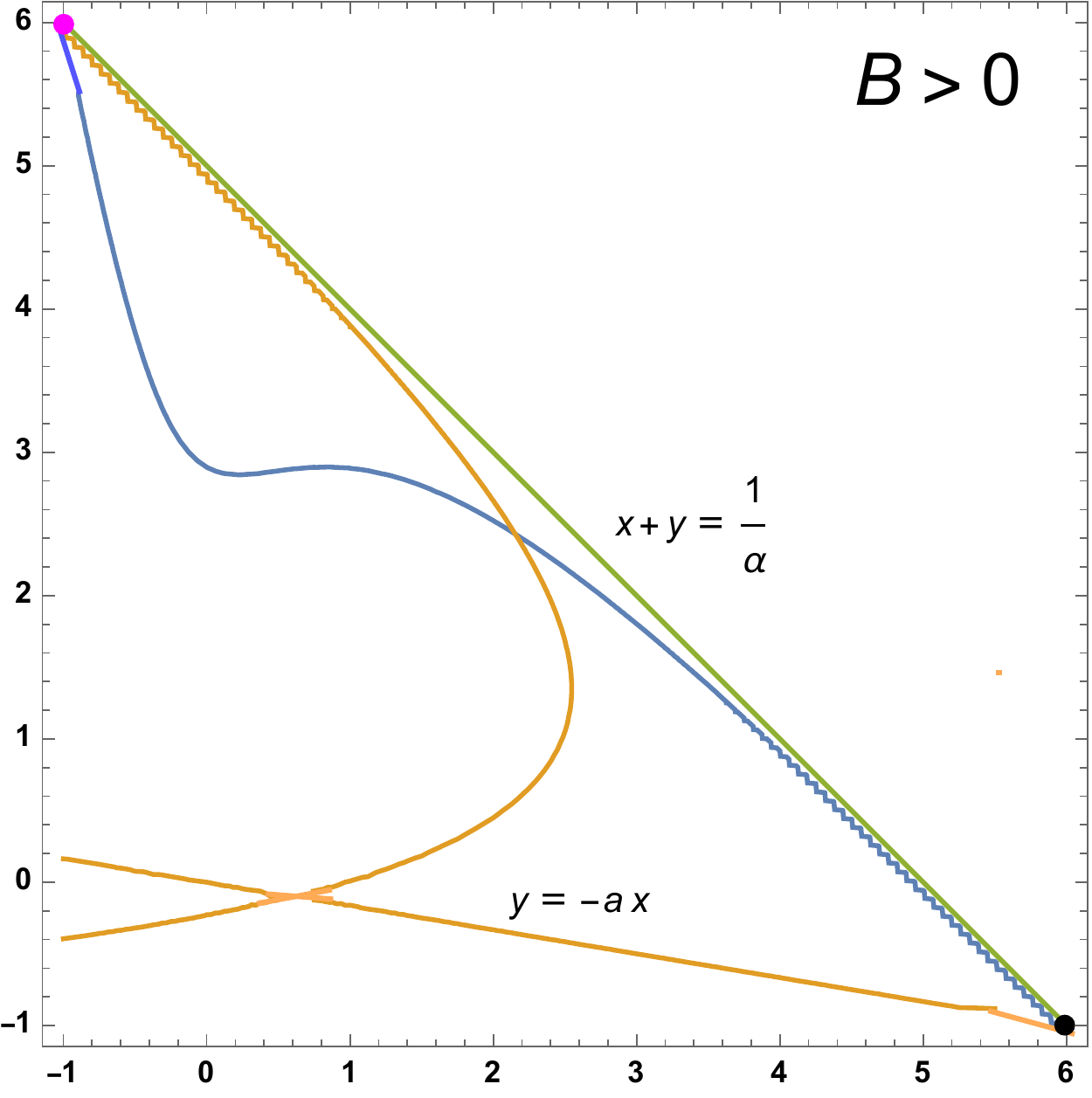}\hskip .2 cm \includegraphics[height= 5.2 cm, angle=0]{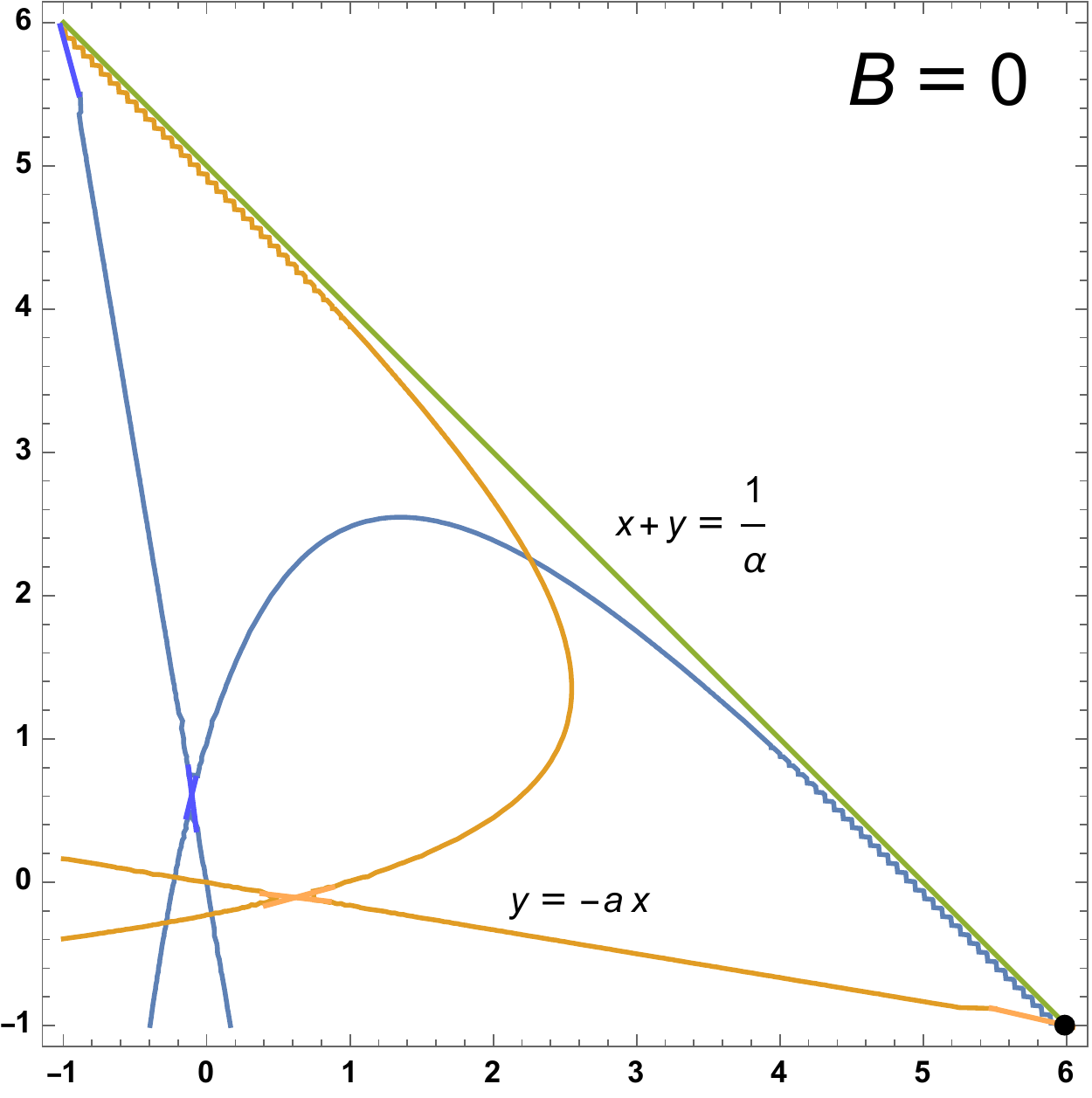}
\hskip .2 cm \includegraphics[height= 5.2 cm, angle=0]{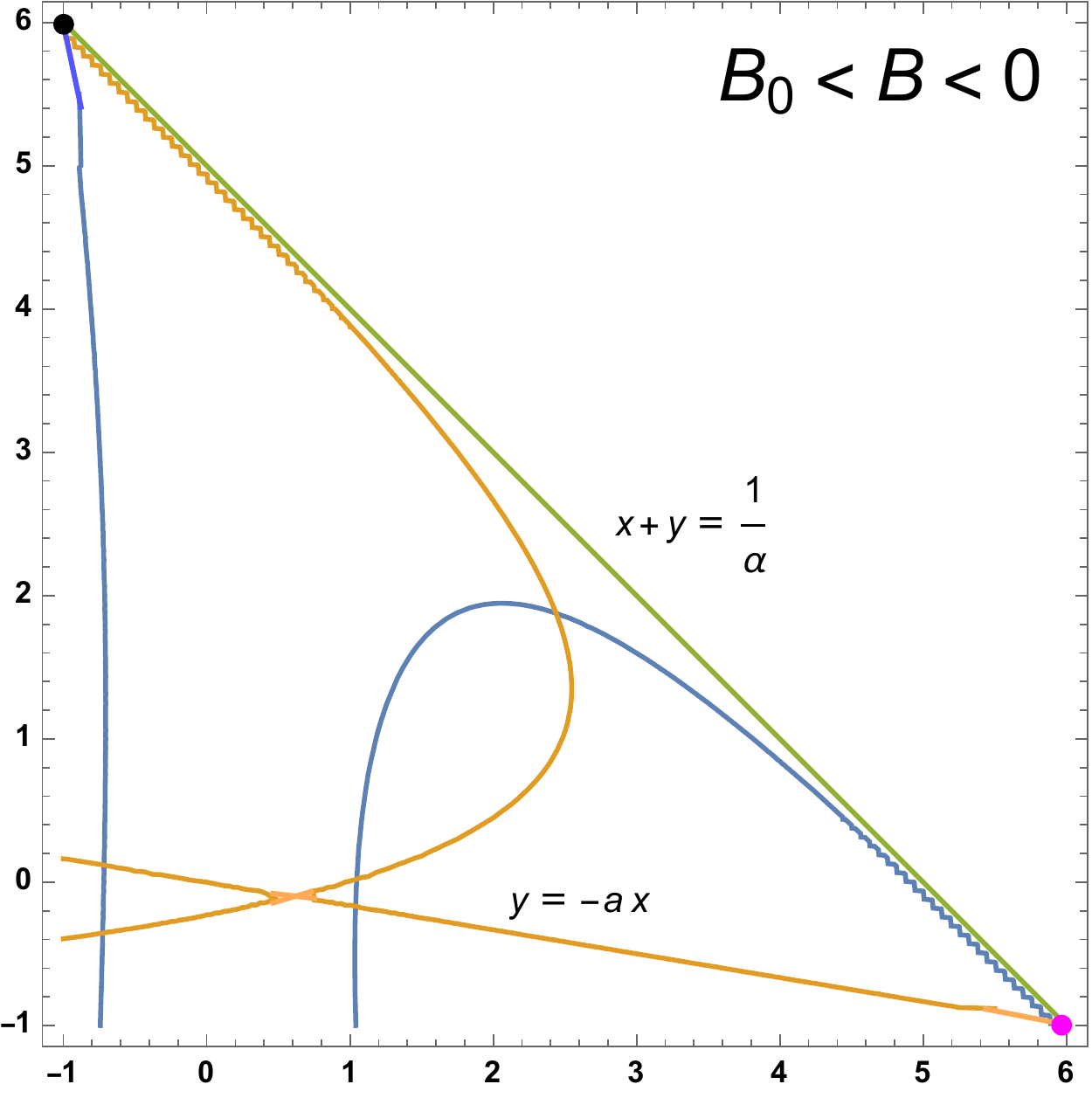} \vskip 0.4cm
\hskip .2 cm\includegraphics[height= 5.2 cm, angle=0]{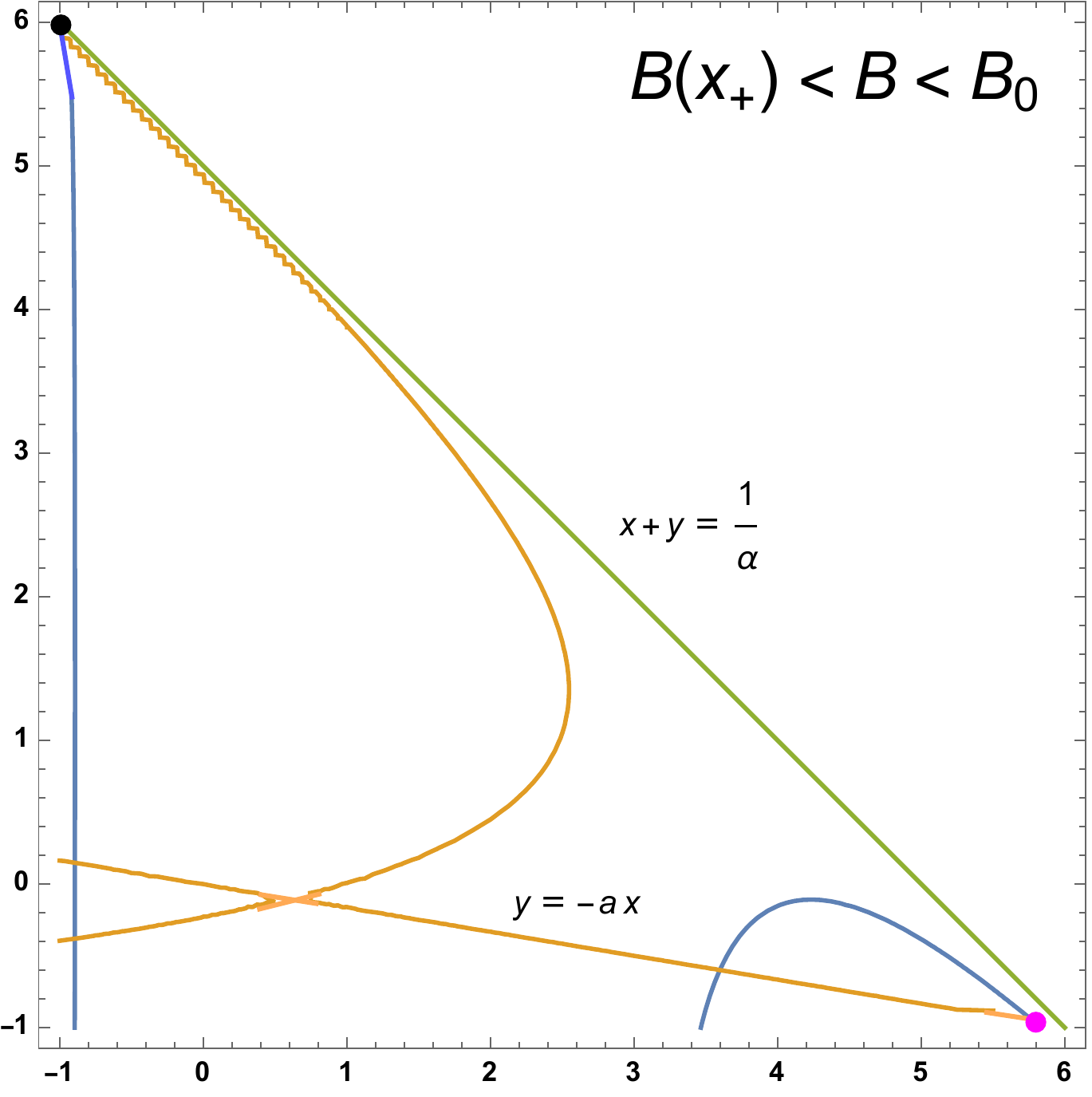}\hskip 1.5 cm \includegraphics[height= 5.2 cm, angle=0]{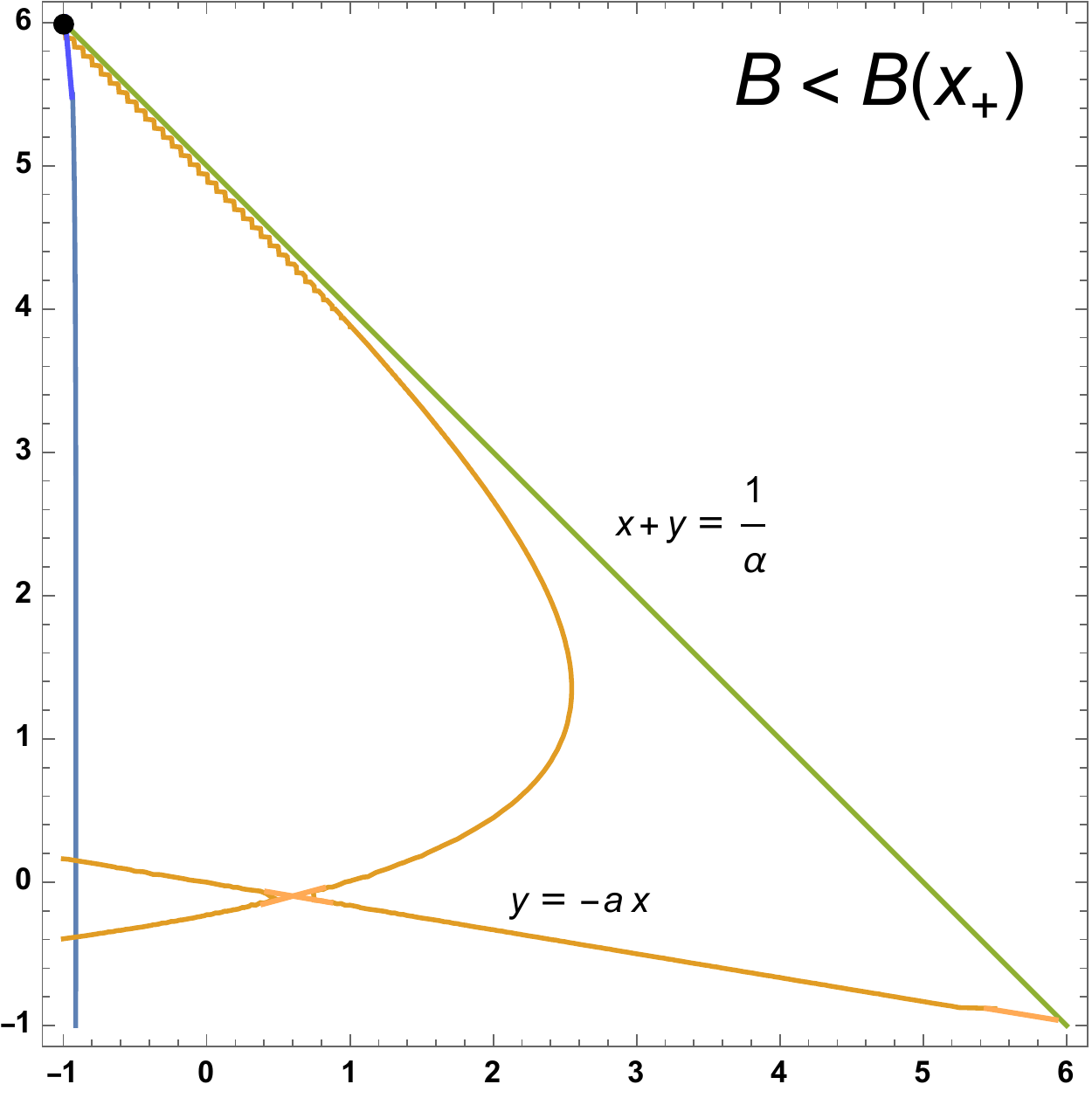}
\end{center}
\vskip -.6 cm
\caption{\small{Typical contour plots for $T<T_0$ in the $x\!-\!y$ plane of the two eqs. in \eqn{lnxy1},
in blue (1st eq.) and orange (2nd eq., for which one of the branches is a straight line).
The intersection points of the blue curves with the orange curves represent solutions of \eqn{lnxy1}.
Stable and metastable solutions are indicated by black and magenta colored dots, respectively.
$B_0$ is the value of $B$ for which the central bulges of the curves would touch (bet$.$ 3rd and 4th plot).
Plots are for $N=7$ and $T=0.9$, and for $B=0.2, 0, -0.4, -2, -3$ 
(note that $B_0\simeq -1.01$ and $B(x_+)\simeq -2.62$) in units of $T_0$.
}}
\label{2-rowsTB2}
\end{figure}

\item
\underline{$B=0$:} 
The system is symmetric under $x \leftrightarrow y$ and we recover the known one-row solution on the
$y=-ax$ line for $x>0$. 
There is also a stable solution on the $y=-x/a$ line for $x<0$, which is equivalent to the previous one,
representing spontaneous magnetization in direction $x_2$.

\item
\underline{$B(x_+)<B<0$:} 
We recover the known one-row solution on the $y=-a x$ line for $x>0$, but now it is metastable.
There is a stable two-row solution with $x<0, y>0$.

\item
\underline{$B<B(x_+)$:} 
There are no stable one-row solutions, in accordance with region $D$ of the one-row phase
diagram of fig. \ref{TBphase}. There is a unique stable two-row solution.

\end{itemize}

\no
We note that there are no solutions with $x=y$  since this is not a consistent truncation of the system \eqn{lnxy1}, unless  $B= 0$ 
in which case we already know that the corresponding two-row solution is unstable.  Thus the symmetry breaking pattern
$SU(N) \to SU(N-2) \times SU(2) \times U(1)$ is never realized.

Recalling fig. \eqn{TBphase},
the picture that emerges, at least for $T<T_0$, is that a two-row solution coexists with the
one-row solution in both regions $C_1 , C_2$. The two-row solution is metastable in
$C_2$ ($B>0$) and becomes
absolutely stable in $C_1$ ($B<0$). The one-row solution is absolutely stable in $C_2$,
becomes metastable in $C_1$, and ceases to
exist in region $D$, leaving the two-row state as the only stable solution there. The line $B=0$ is a
metastability frontier between one-row and two-row solutions for $T<T_0$. We expect this picture to
extend for a range of temperatures $T>T_0$, with a two-row state coexisting with the one-row one
outside of region $D$, although for high enough temperatures the two-row solution should cease
to exist.

\section{Conclusions}

The thermodynamic properties and phase structure of the $SU(N)$ ferromagnet emerge as surprisingly
rich and nontrivial, manifesting qualitatively new features compared to the standard $SU(2)$ ferromagnet.
The phase structure of the system, in particular, is especially rich and displays various phase transitions.
Specifically, at zero magnetic field the system has three critical temperatures (vs. only one for $SU(2)$),
one of them signaling a crossover between two metastable states. Spontaneous breaking of the global
$SU(N)$ group in the ferromagnetic phase at zero external magnetic fields happens only in the
$SU(N) \to SU(N-1) \times U(1)$ channel. In the presence of a nonabelian magnetic field with $M$
nontrivial components ($M<N$), the explicit symmetry breaking (paramagnetic state) is
$SU(N) \to SU(N-M) \times U(1)^M$, while the spontaneous breaking (ferromagnetic state)
is $SU(N) \to SU(N-M-1) \times U(1)^{M+1}$. Finally, due to the presence of metastable states,
the system exhibits hysteresis phenomena both in the magnetic field and in the temperature.

The model studied in this work, and its various generalizations described below, could be relevant in a
variety of physical situations. It could serve as a phenomenological model for physical ferromagnets, in
which the interaction between atoms is not purely of the dipole type and additional states participate in
the dynamics. In such cases, the $SU(N)$ interactions could appear as perturbations on top
of the $SU(2)$ dipole interactions, leading to modified thermodynamics. The model could also be
relevant to the physics of the quark-gluon plasma \cite{QGP}, which can be described as a fluid of
particles carrying $SU(3)$ degrees of freedom, assuming their $SU(3)$ states interact.
Exotic applications, such as matrix models and brane models in string theory, can also be envisaged
(see, e.g. \cite{Mar,Tur}).

Various possible generalizations of the model, relevant to or motivated by potential applications,
and related directions for further investigation suggest themselves. They can be
organized along various distinct themes: starting with atoms carrying a higher representation of $SU(N)$,
generalizing the form of the two-atom interaction, or including three- and higher-atom interactions.

The choice of fundamental representations for each atom was imposed by the physical requirement of
invariance of their interaction under common change of basis for the atom states. Its effect on the
thermodynamics
is to "bias" the properties towards states with a large fundamental content. This manifests, e.g., in the
qualitatively different properties of the system under positive and negative magnetic fields (with respect
to the system's spontaneous magnetization). Starting with atoms carrying a higher irrep of $SU(N)$ would
modify these properties. In particular, starting with atoms in the adjoint of $SU(N)$ would eliminate this
bias altogether. It might also eliminate phases of spontaneous magnetization, and this is worth investigating.

The interaction of atoms $j_{r,a} j_{s,a}$ was isotropic in the group indices $a$, an implication of the
requirement of invariance under change of basis. Anisotropic generalizations of the form \eqn{hab} can
also be considered, involving an "inertia tensor" $h_{ab}$ in the group. Clearly this generalization
contains the higher representation generalizations of the previous paragraph as special cases.
E.g., $SU(2)$ interactions with the atoms in
spin-1 states can be equivalently written as $SU(3)$ fundamental atoms with a tensor $h_{ab}$
equal to $\delta_{ab}$ when $a,b$ are in the $SU(2)$ subgroup of $SU(3)$ that admits the fundamental
of $SU(3)$ as a spin-1 irrep, and zero otherwise. The more interesting special case in which $h_{ab}$
deviates from $\delta_{ab}$ only along the directions of the Cartan generators, in the presence of
magnetic fields along these directions, seems to be the most motivated and most tractable, and is worth
exploring. The phase properties of the model under generic $h_{ab}$ is also an interesting issue.

Including higher than two-body interactions between the atoms is another avenue for generalizations.
Physically, such terms would arise from higher orders in the perturbation expansion of atom interactions,
and would thus be of subleading magnitude, but the possibility to include them is present.
Insisting on invariance
under common change of basis and a mean-field approximation would imply that such interactions
appear as higher Casimirs of the global $SU(N)$ and/or as higher powers of Casimirs, the most general
interaction being a general function $f(C^{(2)}, \cdots, C^{(N-1)})$ of the full set of Casimirs of the
global $SU(N)$. These can be readily examined using the formulation in this paper and may lead to
models with richer phase structure. An interesting extension of this study is in the context of
topological phases nonabelian models. Such topological phases have been proposed in one dimension \cite{RQ,CFLT,TLC,RPAR} and it would be interesting to see if they exist in higher dimensions.

Another independent direction of investigation is the large-$N$ limit of the model. This could be 
conceivably relevant to condensed matter situations involving interacting Bose condensates, or to
more exotic situations in string theory and quantum gravity. The presence of two large parameters,
$n$ and $N$, presents the possibility of different scaling limits. These will be explored in an upcoming
publication.

Finally, the nontrivial and novel features of this system offer a wide arena for experimental verification and
suggest a rich set of possible experiments. The experimental realization of this model, or the demonstration
of its relevance to existing systems, remain as the most interesting and physically relevant open issues.

\subsection*{Acknowledgements}

We would like to thank I. Bars for a very useful correspondence, D. Zoakos for help with numerics,
and the anonymous Reviewer for comments and suggestions that helped improve the manuscript.\\
A.P. would like to thank the Physics Department of the U. of Athens for hospitality during the initial stages of this work.
His visit was financially supported by a Greek Diaspora Fellowship Program (GDFP) Fellowship.\\
The research of A.P. was supported in part by the National Science Foundation
under grant NSF-PHY-2112729 and  by PSC-CUNY grants 65109-00 53 and 6D136-00 02.\\
The research of K.S. was supported by the Hellenic Foundation for
Research and Innovation (H.F.R.I.) under the ``First Call for H.F.R.I.
Research Projects to support Faculty members and Researchers and
the procurement of high-cost research equipment grant'' (MIS 1857, Project Number: 16519).

\appendix

\section{Analysis of one-row states with a magnetic field}

In this appendix we present the details of the analysis that we have summarized in the main text.

\begin{figure} [h!] 
\begin{center}
\includegraphics[height= 4.5cm, angle=0]{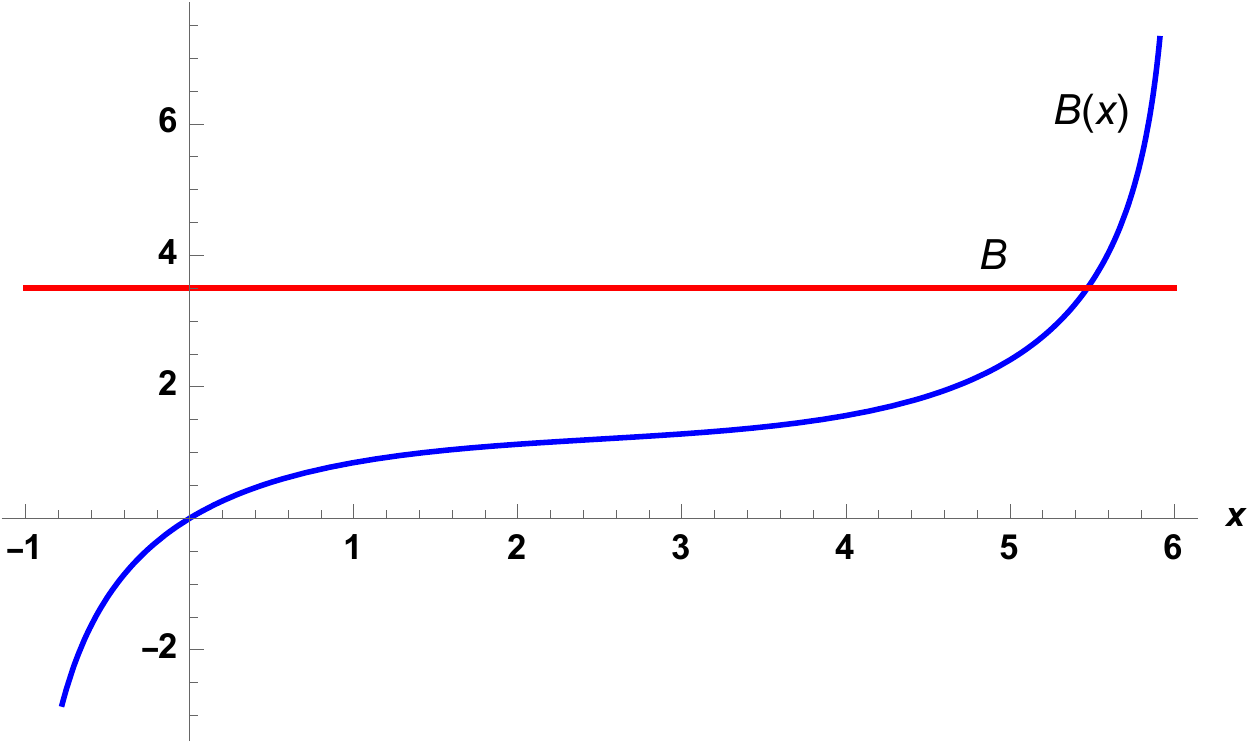}
\end{center}
\vskip -.7 cm
\caption{\small{Plot of  $B(x)$  for  $T > T_+$.}} 
\label{BTgTp}
\end{figure}

\no
\underline{$T>T_+$} (fig. \ref{BTgTp}): $B(x)$ is increasing and all values of $x$ are stable, so
there exsits a unique stable solution for all $B$, one-row for $B>0$ and its conjugate for
$B<0$. 

\vskip -.3 cm

\begin{figure} [th!] 
\begin{center}
\includegraphics[height= 4 cm, angle=0]{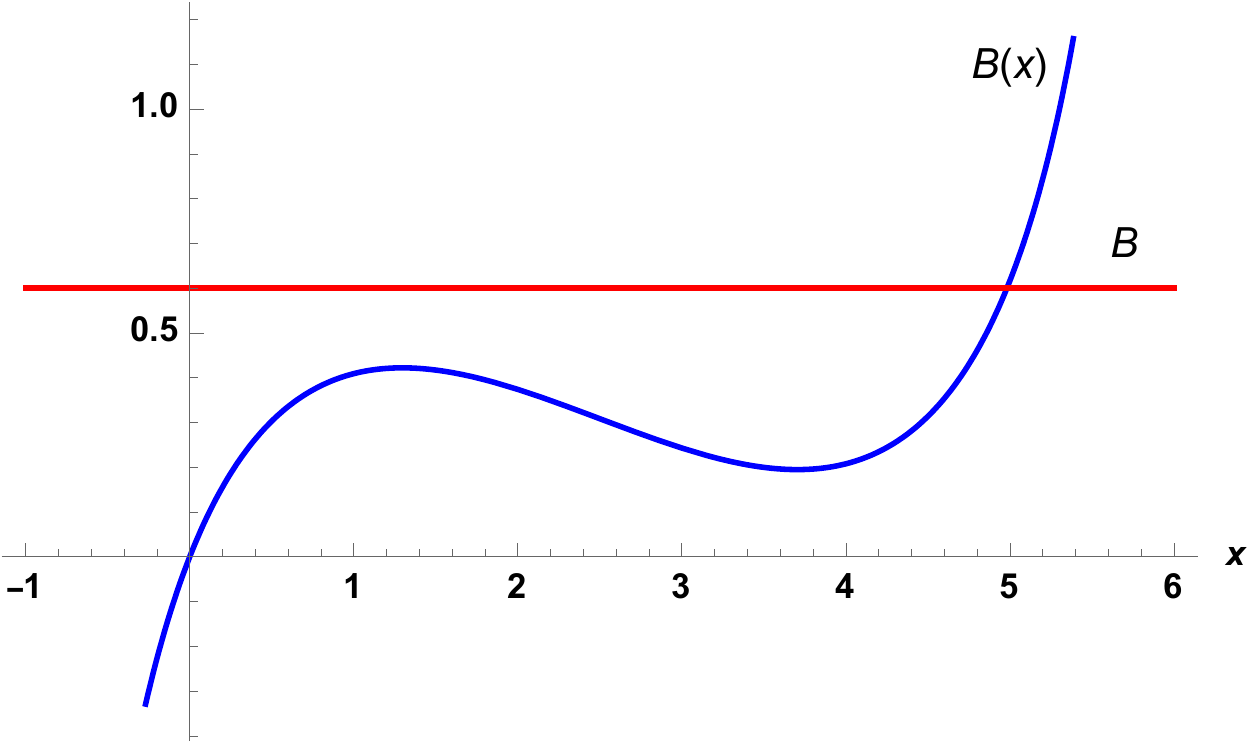}\hskip .3 cm
\includegraphics[height= 4 cm, angle=0]{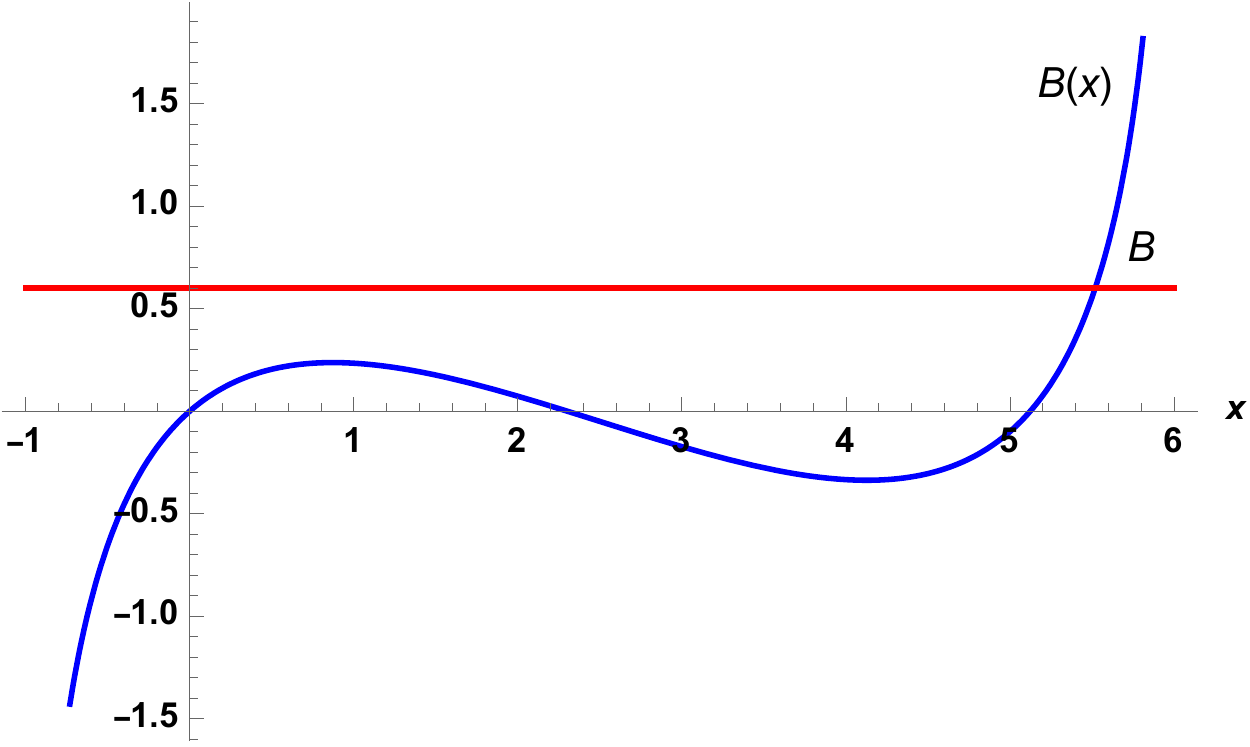}
\end{center}
\vskip -.6 cm
\caption{\small{Plot of  $B(x)$  for  $T_c<T < T_+$ (Left) and for  $T_-<T < T_c$ (Right). }}
\label{BTcTpTn}
\end{figure}

\no
\underline{$T_- <T<T_+$} (fig. \ref{BTcTpTn}): We have stability for $-1<x<x_-$ and $x_+ <x <N-1$,
the regions of increasing $B(x)$. Further, note that $x_->0$ and $B(x_-) > B(x_+)$, and that $B(x_-)>0$
while $B(x_+)$ can be positive or negative. The critical temperature $T_c$ satisfies the condition
\be
T = T_c\quad  \Longleftrightarrow \quad B(x_+ ) = 0\ ,
\ee
which is precisely the condition \eqn{hgh2}. Note that, for $N>4$, $T_-<T_c<T_+$.
For $T>T_c$, $B(x_+)>0$ and for $T<T_c$, $B(x_+)<0$.

\vskip -0 cm
\no
So within this temperature range we distinguish two sub-cases: 

\no
\underline{$T_c<T<T_+$} (left plot in fig. \ref{BTcTpTn}): we have for various values of the magnetic field:
 
\begin{itemize}

\item
 For $B< 0$, $x$ varies from $-1$ to $0$ and there is a unique stable solution corresponding to the
conjugate one-row YT.

\item
For $0<B<B(x_+)$, $x$ varies from $0$ to some $x<x_-$ and there is a unique stable solution 
corresponding to a one-row YT. 

\item
 For $B(x_+) <B< B(x_-)$ we have two locally stable solutions, one for some $0<x<x_-$ and one for
some $x>x_+$ (a third solution in between is unstable). 
The first one corresponds to an unbroken phase, as it represents a continuous deformation of the
singlet for $B=0$, and the second one to a broken phase.
One is absolutely stable and the other metastable. To decide which, we need to compare their free
energies.

\item
For $B>B(x_-)$, $x$ varies from some value greater than $x_+$ to $N-1$ and there is a unique stable
solution.

\no
Note that for $B=0$ there is only the solution $x=0$, as expected.

\end{itemize}

\no
\underline{$T_-<T<T_c$} (right plot in fig. \ref{BTcTpTn}): we have for various values of the magnetic field:

\begin{itemize}

\item
For $B< B(x_+)$, $x$ varies from $-1$ to some negative value $x$ obtained from $B(x_+)= B(x)$ and
there is a unique stable solution corresponding to the conjugate one-row YT.

\item
 For $B(x_+) <B< B(x_-)$ we have two locally stable solutions, one for some $x<x_-$ and one for some
$x>x_+$. The first one represents an unbroken phase and  the second one a broken phase, as they map
to the singlet and the one-row solutions for $B=0$. One is absolutely stable and the other metastable. 
To decide, we need to compare their free energies.

\item
For $B>B(x_-)$, $x$ varies from some $x> x_+$ to $N-1$ and there is a unique stable solution
corresponding to a one-row YT. 

Note that $B=0$ is in the range $B(x_+ ) <B <B(x_- )$ and we recover
the expected two solutions, $x=0$ (singlet) and $x>0$ (one-row).

\end{itemize}

\begin{figure} [th!] 
\begin{center}
\includegraphics[height= 5 cm, angle=0]{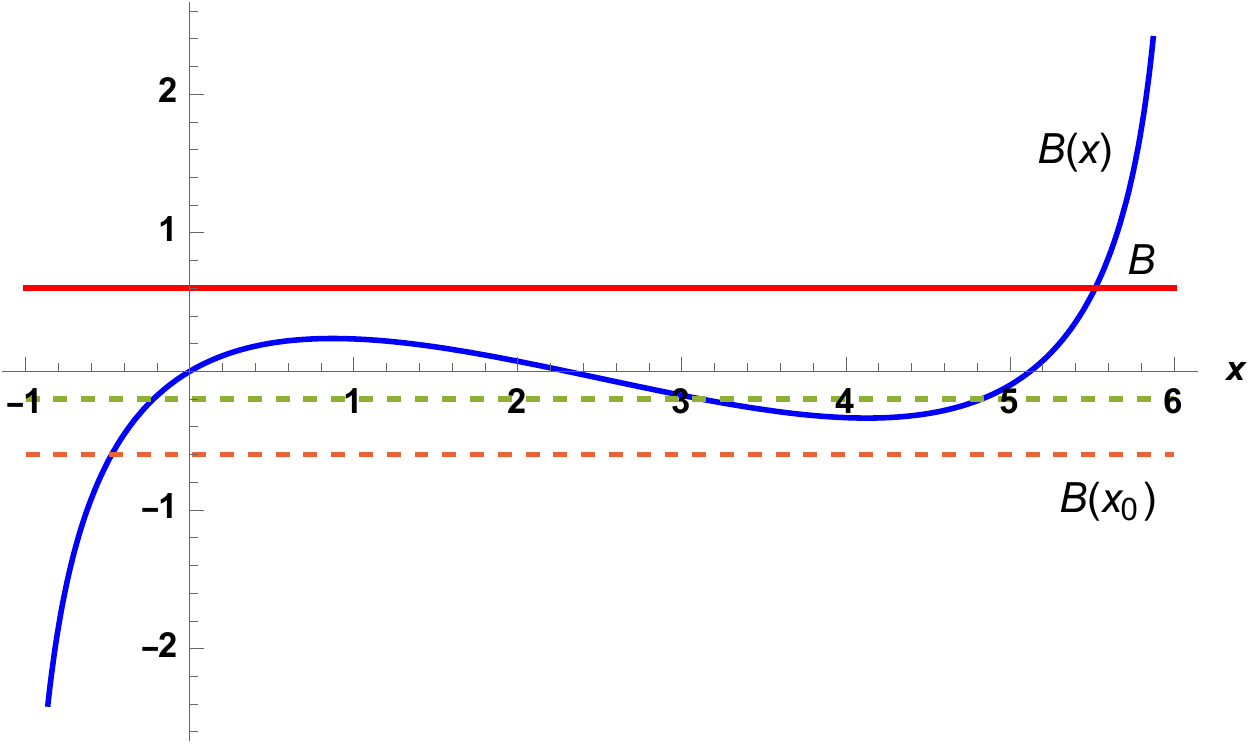}
\end{center}
\vskip -.6 cm
\caption{\small{Plot of  $B(x)$  for  $T_0<T < T_-$. The dashed lines refer to the value $B(x_0 )$
which could be higher (green) or lower (red) than $B(x_+)$. }}
\label{BT0Tm}
\end{figure}

\no
\underline{$T_0<T<T_-$} (fig. \ref{BT0Tm}): The situation is as in case $T_-<T<T_c$, except now
$x$ cannot be more negative than $x_0 = (N-1)(1-T/T_0)$ defined in \eqn{fg9}.  The relative size
of $B(x_0)$ and $B(x_+)$ will also play a role:

\begin{itemize}
\item
For $B < \min\{B(x_0),B(x_+)\}$ there is {\it no} stable one-row solution and the stable solution must
necessarily have more rows. According to the general stability analysis, it must be one with $N-1$ rows
out of which $N-2$ have equal length.

\item
For $\min\{B(x_0),B(x_+)\} <B< \max\{B(x_0),B(x_+)\}$ there is one stable solution, for $x<0$
($x>x_+$) if $B(x_0)<B(x_+)$ ($B(x_0)>B(x_+)$).

\item
For $\max\{B(x_0),B(x_+)\}<B<B(x_-)$ there are two stable solutions, one absolutely stable and
the other metastable. To decide, we need to compare their free energy.

\item
For $B>B(x_-)$, $x$ there is a single stable solution varying from some $x> x_+$ to $N-1$.

\end{itemize}

\vskip -0.5cm

\no
\underline{$T<T_0$}: Only $x>x_+$ solutions are stable. 

\begin{itemize}

\item
For $B<B(x_+)$ there is {\it no} stable one-row solution, and the solution must again be one with
$(N-1)$-rows.
\item
For $B>B(x_+)$ there is one stable solution for $x>x_+$

\end{itemize}

\subsection{Resolving metastability}

To determine which configuration is absolutely stable and which is metastable when there are two
locally stable solutions, we need to compare their free energies.
The free energy of the system is given by \eqn{onsh} with the addition of the magnetic field term,
\be
\begin{split}
\label{onsh11a}
 F_{\rm sym}(x,T)= & {T\ov 1+a}\big(a(1+x)\ln (1+x) +(1-a x)\ln(1-a x)\big)
 \\
 &-{a\ov 2} T_0 x^2 -{B(x)\ov N} x - T \ln N\ ,
\end{split}
\ee
where $a=1/(N-1)$ and where $B=B(x)$ is expressed in terms of $x,T$ via \eqn{gfuiy21}.

\no
To facilitate the comparison, define the modified free energy $\Phi$ (we use $F(x)$ instead of $F_{\rm sym}(x,T)$ for
notational convenience)
\be
{\Phi} (x) = F(x) + {N-2 \over 2N} B(x) \ .
\ee
We can show that ${\Phi}(x)$ and $B(x)$ satisfy
\be
\label{FN2}
{\Phi}(N-2-x) = {\Phi}(x) 
\ee
and
\be
B(x) + B(N-2-x) = 2 T\, { \ln(N-1)} - T_0{N(N-2) \over N-1}\ .
 \label{Bex}
 \ee
For two solutions with different $x, x'$ but the same $B$, $F(x) - F(x' ) = {\Phi} (x)-{\Phi} (x' )$,
so we can compare their $\Phi$ to resolve metastability. 
At the transition point, when the two
solutions have the same free energy, \eqn{FN2} implies
\be
{\Phi} (x) = {\Phi} (x')\quad \Longrightarrow \quad x' = N-2 -x
\ee
and from (\ref{Bex}) with $B(x) = B(x' ) = B(N-2-x)$ the magnetic field $B_t$ at which this happens is
\be
B_t = T\,  \ln(N-1) -{T_0\ov 2} {N(N-2) \over N-1}\ .
\ee
For fixed $B$, this gives the transition temperature $T_t$ at which the two solutions will transit from
stable to metastable as
\be
T_t = { 1 \over \ln (N-1)} \left( B +T_0\, {N(N-2) \over 2(N-1)} \right)\ .
\ee
For $B=0$ this reproduces the temperature $T_1$ and magnetization $x_1$ that we determined
before in \eqn{tempt1}.

\section{Analysis of two-row states in a magnetic field}

In this appendix we investigate in detail two-row solutions, including their $(N-1)$-row conjugates.
We analyze the conditions for their stability, present the corresponding YT of their irreps, and derive
numerical results for the case of temperatures $T < T_0$.

\no
To proceed, we write the coefficients $C_i$ defined in \eqn{2var} in
terms of the variables  $x$ and $y$ of \eqn{xydefs}. We obtain
\be
\begin{split}
& C_1^{-1} =N \, A(x)\ ,\qquad C_2^{-1}= N \, A(y)\ ,
\\
&C_i^{-1} = N \, A\big(-\a (x+y)\big)\ ,    \qquad i =3,4,\dots, N\ ,
\end{split}
\ee
with our usual $A(x)$ defined in \eqn{aax}.
The variables $x$ and $y$ are restricted by the conditions $0<x_i <1$ to the range
\be
x,y>-1\ ,\qq x+y<{1\ov \a} =N-2\ .
\ee
Thus, the allowed solutions are within the triangle in the $(x,y)$-plane with corners at the
points $(1/a,-1)$, $(-1,1/a)$ and $(-1,-1)$, depicted in fig. \ref{2rows}.
This triangle is further subdivided by the curves $y=x$, $y=-a x$ and $y=-x/a$ into six regions
representing the possible ordering of $x_1, x_2, x_i$ ($i\geqslant 3$) and thus the various YT renditions
of the two-row solution. These regions are accordingly labelled by $(ijk)$ for $x_i>x_j>x_k$.
Assuming $N>3$, most of the $C_i$'s are proportional to $1/A(-\a(x+y))$, so we choose solutions with 
$A(-\a(x+y))>0$. Then at most one of the functions $A(x)$ and $A(y)$ can be negative.
We list below the various possibilities together with conditions for stability:

{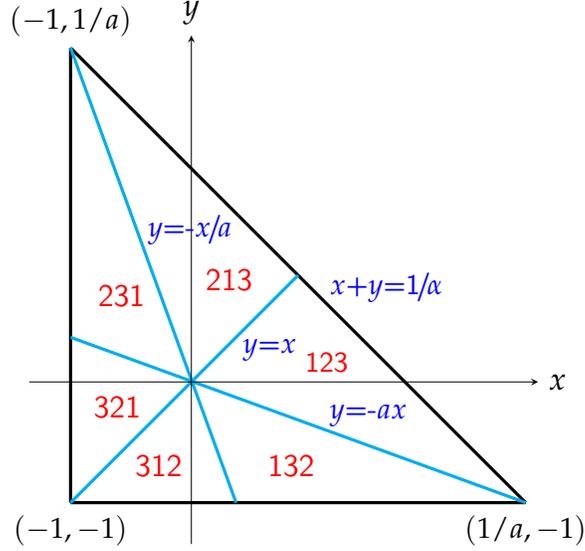
\begin{figure} [th!] 
\vskip-0.5cm
\centering 
{
\begin{tikzpicture}[scale=0.55]

\tikzset{big arrow/.style={decoration={markings,mark=at position 1 with {\arrow[scale=3,#1,>=stealth]{>}}},postaction={decorate},},big arrow/.default=black}

\draw[very thick,-] (-1,-1) -- (10,-1)node[below]{\small$(1/a,-1)$};
\draw[very thick,-] (-1,-1)node[below]{\small$(-1,-1)$} -- (-1,10)node[above]{\small$(-1,1/a)$};
\draw[very thick,-] (-1,10) -- (10,-1);
\draw[cyan,very thick,-] (-1,-1) -- (4.5,4.5);
\draw[cyan,very thick,-] (10,-1) -- (-1,3);
\draw[cyan,very thick,-] (-1,10) -- (3,-1);

\draw [very thin,-stealth](-2,1.92){--++(12.3,0)}node[right] {\color{black} $x$};
\draw [very thin,-stealth]((1.92,-2){--++(0,12.3)}node[above] {\color{black} $y$};

\draw[-](3.5,-0.1)node[right]{\color{red} {\textsf{132}}};
\draw[-](4.4,2.4)node[right]{\color{red} {\small\textsf{123}}};
\draw[-](2,4.4)node[right]{\color{red} {\textsf{213}}};
\draw[-](-0.6,4)node[right]{\color{red} {\textsf{231}}};
\draw[-](-0.7,1.3)node[right]{\color{red} {\textsf{321}}};
\draw[-](0.3,-0.1)node[right]{\color{red} {\textsf{312}}};
\draw[-](0.6,5.6)node[right]{\color{blue} {\small \textsf{$y$=-$x$\hskip-0.05cm/\hskip-0.05cm $a$}}};
\draw[-](5,1.1)node[right]{\color{blue} {\small \textsf{$y$=-$a$$x$}}};
\draw[-](2.9,2.7)node[right]{\color{blue} {\small \textsf{$y$=$x$}}};
\draw[-](5,4.2)node[right]{\color{blue} {\small\textsf{$x$+$y$=1\hskip-0.05cm/\hskip-0.05cm $\alpha$}}};
\end{tikzpicture}
\vskip -.3 cm
\caption{\small{The domain of $x,y$. Coordinate axes do not create subdivisions.  }}
\label{2rows}
}
\end{figure}}

\no
\underline{Region A,  or $(123)$:}   \hskip .3 cm $x_1>x_2>x_3$, with
 \be
 \label{regA1}
-a x<y< x\quad \Longrightarrow\quad  A(x)<A(y)<A(-\a(x+y))\ .
\ee
The stability condition is
\be
 \label{regA2}
A(y)>0\ ,\qq \a [A(x)+ A(y)] A(-\a(x+y)) + A(x) A(y) >0 \  .
\ee
In this region all $C_i$'s are positive, excect $C_1$ which may have either sign. 
We define 
 \be
  \label{regA3}
\begin{split}
 \ell_1 = {(1+\a)x +\a y\ov N}\ ,\qquad \ell_2 = {(1+\a)y +\a x\ov N}\ , \qquad \ell_1>\ell_2>0\ .
\end{split} 
\ee
The YT has the partition
\be
\label{partA}
(\ell_1,\ell_2)\ .
\ee
Hence, it represents a two-row YT with $\ell_1$ and $\ell_2$ boxes, respectively. 
This follows from the fact that $x_3$ is the smallest among the three $x_i$'s and appears $N-2$ times.  

\no
\underline{Region B,  or $(132)$:}  \hskip .3 cm $x_1>x_3>x_2$, with 
\be
  -x/a <y<-a x\ ,\qq A(x)<A(-\a(x+y))<A(y)\ . 
\ee
The stability condition is
\be
 A(-\a(x+y)>0\ ,\qq \a [A(x)+ A(y)] A(-\a(x+y)) + A(x) A(y) >0 \  .
\ee
In this region all $C_i$'s are positive expect $C_1$ which may have either sign. 
We define 
\be
\ell_1 = {x - y\ov N}\ ,\qquad \ell_2 = -{(1+\a)y +\a x\ov N}\ ,  \qquad \ell_1>\ell_2>0\ .
\ee
The $(N-1)$-row YT has the partition
\be
\label{partB}
(\ell_1,\underbrace{\ell_2,\ell_2,\dots, \ell_2}_{N-2})\ .
\ee
Hence, it represents  a YT with $\ell_1$ boxes in the first line and $\ell_2$ boxes in the following $N-2$ lines.
This follows from the fact that the smallest among the three $x_i$'s,  $x_2$ appears only once.  

\no
\underline{Region C,  or $(213)$:} \hskip .3 cm $ x_2>x_1>x_3$, with
\be
-a y<x< y\ ,\qq A(y)<A(x)<A(-\a(x+y))\ .
\ee
The stability condition is
\be
A(x)>0\ ,\qq \a [A(x)+ A(y)] A(-\a(x+y)) + A(x) A(y) >0 \  .
\ee
In this region all $C_i$'s are positive expect $C_2$ which could be either positive or negative.  
Defining
\be
 \ell_1 = {(1+\a)y +\a x\ov N}\ ,\qquad \ell_2 = {(1+\a)x +\a y\ov N}\ ,  \qquad \ell_1>\ell_2>0\ ,
\ee
this case represents a two-row YT  with the partition \eqn{partA}.

\no
\underline{Region D,  or $(231)$:}  \hskip .3 cm  $ x_2>x_3>x_1$, with
\be
-a x<y<- x/a\ ,\qq A(y)<A(-\a(x+y))<A(x)\ .
\ee
The stability condition is
\be
 A(-\a(x+y)>0\ \ ,\qq \a [A(x)+ A(y)] A(-\a(x+y)) + A(x) A(y) >0 \  .
\ee
In this region all $C_i$'s are positive expect $C_2$ which could be either positive or negative.
Defining
\be
\ell_1 = {y-x \ov N}\ ,\qquad \ell_2 = -{(1+\a)x +\a y\ov N}\ ,  \qquad \ell_1>\ell_2>0\ .
\ee
this case represents a YT  with the partition \eqn{partB}.

\no
\underline{Region E,  or $(312)$:}  \hskip .3 cm  $ x_3>x_1>x_2$, with $ y<x<-a y $, equivalently $ x_3>x_1>x_2$, with
\be
 y<x<-a y \ ,\qq A(-\a(x+y))<A(x)<A(y)\ .
\ee
 The stability condition is
\be
A(-\a(x+y))>0\ .
\ee
In these two regions all $C_i$'s are positive.
Defining
\be
 \ell_1 =-{(1+\a)y +\a x\ov N}\ ,\qquad \ell_2 = {x-y\ov N}\ ,  \qquad \ell_1>\ell_2>0\ ,
\ee
this case represents a $(N-1)$-row  YT  with the partition 
\be
\label{partE}
(\underbrace{\ell_1,\ell_1,\dots, \ell_1}_{N-2},\ell_2)\ ,
\ee

\no
\underline{Region F,  or $(321)$:}  \hskip .3 cm $  x_3>x_2>x_1$, with
\be
x<y<-a x\,\qq A(-\a(x+y))<A(y)<A(x)\ .
\ee
The stability condition is
\be
A(-\a(x+y))>0\ .
\ee
In these two regions all $C_i$'s are positive.
Defining
\be
  \ell_1 =-{(1+\a)x +\a y\ov N}\ ,\qquad \ell_2 = {y-x\ov N}\ , \qquad \ell_1>\ell_2>0\ .
\ee
this case represents a $(N-1)$-row  YT  with the partition \eqn{partE}.

\subsection{The case of low temperatures}

We consider temperatures $T<T_0$ for which spontaneous magnetization exists. 
Various cases arise depending on the sign of $B$ and, if negative, on the value 
$B(x_+)<0$ with $x_+$ defined in \eqn{xpm0} (recall that $B(x_+)$ is negative for 
$T<T_0$) and some other intermediate value $B_0$ to be defined shortly.
In the plots below all points on the red line $y=-ax$ solve identically the second equation in the system \eqn{lnxy1}, whereas the first one reduces 
to the equation for the one-row configuration \eqn{gfuiy11}.
This line can be approached for $x>0$ from the regions A and B and for $x<0$ from the regions D and F.
In addition, the stability conditions for these regions reduce to those in \eqn{jjjs1}. All cases below map to
one of the plots in Fig. \ref{2-rowsTB2}.

\begin{itemize}

\item
\underline{$B>0$:} We know that the one-row configuration 
has a stable solution which in this two-parameter plot is on the $y=-ax$ red line for $x>0$. The intersection point in the middle (region C in 
Fig. \ref{2rows}) is unstable, whereas that on the upper left corner is, having a higher value for the free energy, metastable (region C). 

\item
\underline{$B=0$:} We have included the case with $B=0$, which is 
symmetric with respect to the $y=x$ line. We know that the one-row configuration 
has a stable solution which in this two-parameter plot is on the $y=-ax$ red line for $x>0$. 
Due to the above symmetry there is stable solution also on the $y=-x/a$ line for $x<0$, which however is equivalent to the above.
The other three intersecting points are unstable.

\item
\underline{$B_0<B<0$:} The value of the magnetic field $B_0$ arises 
when the two curves in the plot meet tangentially. 
The one-row configuration has a stable solution which in this two-parameter plot is on the $y=-ax$
red line for $x>0$. 
However, this becomes now metastable, as the stable intersection point is on the upper left corner
(region D in fig. \ref{2rows}), corresponding to an $(N-1)$-row YT with partition \eqn{partB}.
The other intersection points are unstable.

\item
\underline{$B(x_+)<B<B_0$:} There are typically 
three intersection points along the $y=-a x$ line corresponding to the one-row configuration, the far right is now metastable and the
other two unstable. However, there are two additional intersecting  points in the far left of that plot, the lower one being unstable and the upper one  stable (region D in Fig. \ref{2rows}) corresponding to an
$(N-1)$-row YT with partition \eqn{partB}.

\item
\underline{$B<B(x_+)$:} There is  
one intersection  points along the $y=-a x$ line corresponding to the one-row configuration. This has $x<0$ and is, as we have shown unstable.
In fact, there is no stable one-row configurations for $B<B(x_+)$.
Among the other two intersection points the lower one is unstable and the upper one is stable (region D in Fig. \ref{2rows}) 
and again it corresponds to a $(N-1)$-row YT with partition \eqn{partB}.

\end{itemize}


\end{document}